\documentstyle[prd,aps]{revtex}

\newcommand{\bea}{\begin{eqnarray}}
\newcommand{\eea}{  \end{eqnarray}}

\begin{document}

\title{Classical evolution and quantum generation in generalized gravity
       theories including string corrections and tachyon: Unified analyses}
\author{Jai-chan Hwang${}^{(a)}$ and Hyerim Noh${}^{(b)}$ \\
        ${}^{(a)}$ Department of Astronomy and Atmospheric Sciences,
                   Kyungpook National University, Taegu, Korea \\
        ${}^{(b)}$ Korea Astronomy and Space Science Institute, Daejon, Korea \\
        E-mails: ${}^{(a)}$ jchan@knu.ac.kr, ${}^{(b)}$ hr@kao.re.kr
        }
\date{\today}
\maketitle

\begin{abstract}

We present cosmological perturbation theory based on generalized gravity
theories including string theory correction terms and a tachyonic complication.
The classical evolution as well as the quantum generation processes
in these variety of gravity theories are presented in unified forms.
These apply both to the scalar- and tensor-type perturbations.
Analyses are made based on the curvature variable in two different gauge
conditions often used in the literature in Einstein's gravity;
these are the curvature variables in the comoving (or uniform-field) gauge
and the zero-shear gauge.
Applications to generalized slow-roll inflations and its consequent
power spectra are derived in unified forms which include wide range of
inflationary scenarios based on Einstein's gravity and others.

\end{abstract}

\noindent


\section{Introduction}
                                               \label{sec:Introduction}

Cosmological linear perturbation theory \cite{Lifshitz-1946} has central
importance in the current cosmological investigations of the large-scale
structure and the cosmic microwave background radiation.
A rather successful scenario can be made based on Einstein's gravity
with varying use of the diverse (but ordinary) fluids and fields as the
energy-momentum content.
Relativistic gravity theories more general than Einstein's gravity
are ubiquitous in the literature; some are variants of Einstein's
gravity while others are more generalized forms with natural correction
terms arising in the quantum corrections or in the attempt of unified
theories like string/M-theory program.
Thus, it would be interesting to formulate corresponding cosmological
perturbation analyses in these generalized forms of relativistic
gravity theories.
It would be naturally more interesting if we could make a unified
formulation of handling the perturbations in the context of
generalized theories including the Einstein's theory as a case.
This is our purpose of the presentation.
We consider generalized forms of gravity theories expressed as actions
in eqs.
(\ref{MSF-action},\ref{GGT-action},\ref{Tachyon-action},\ref{String-corrections},\ref{String-axion-correction}).

In the literature equations in two different gauge conditions
are popularly used.
In terms of the three-space curvature perturbation $\varphi$
the two gauges are the comoving gauge $v \equiv 0$ (or equivalently
in the minimally coupled scalar field,
the uniform-field gauge $\delta \phi \equiv 0$),
and the zero-shear gauge $\chi \equiv 0$ conditions.
As each of these two gauge conditions fixes the temporal
gauge mode completely the variables are equivalently gauge invariant
and correspond to the combinations $\varphi_v$ (or $\varphi_{\delta \phi}$)
and $\varphi_\chi$ in eq. (\ref{GI-variables}).
In the presence of background curvature we need generalization of $\varphi_v$
which we call $\Phi$ to have the unified form, and
in the case of the generalized gravity we also need generalization of
$\varphi_\chi$ which we call $\Psi$ to have the unified form, see eqs.
(\ref{Phi-def},\ref{Phi-def-MSF},\ref{Psi-def},\ref{Phi-def-Tachyon},\ref{Psi-def-string}).
In all the gravity theories we are considering we can successfully present
the perturbation equations in exactly the same form as in Einstein's gravity.
Thus, we can present the consequent classical evolution and
quantum generation processes in unified forms.
Such unified analyses are practically useful to handle the structure evolution
because
\begin{quote}
``{\sl it allows to handle situations when one type of gravity theory
       switches to another type in the early universe.}''
\end{quote}

In \S \ref{sec:Notation} we present our notation, summary
of the gauge issue, and the fundamental equations in a gauge-ready form.
In \S \ref{sec:two-gauges} we present the closed form equations
using the $\Phi$ and $\Psi$
which are generalizations of $\varphi_v$ and $\varphi_\chi$, respectively,
in the cases of the fluid, field, and generalized forms of gravity theories.
In \S \ref{sec:CE} we present the unified form equations for all
the gravity theories considered in this work and present variety of
exact and asymptotic solutions available.
In \S \ref{sec:QG} we present the quantum generation process in unified
form starting from the action formulation.
An exact form of inflation generated power spectrum is derived under an
assumption of the background which is quite general so that it includes
various inflation scenarios suggested in the literature as cases.
Thus, we present the final inflationary spectra in unified forms
which can be compared with the cosmic microwave background radiation (CMB)
and the large-scale structure observations.
Our classical evolution and the quantum generation processes are presented
for both the scalar-type and the tensor-type perturbations in
unified forms.
In \S \ref{sec:Discussions} we summarize the new discoveries in
our presentation and provide a discussion.

\section{Strategy and Basic Equations}
                                               \label{sec:Notation}

\subsection{Notation}

As the metric we consider the Robertson-Walker spacetime with
the scalar- and tensor-type perturbations
\bea
   & & d s^2 = - a^2 \left( 1 + 2 \alpha \right) d \eta^2
       - 2 a^2 \beta_{,\alpha} d \eta d x^\alpha
       + a^2 \left( g^{(3)}_{\alpha\beta}
       + 2 \varphi g^{(3)}_{\alpha\beta}
       + 2 \gamma_{,\alpha|\beta}
       + 2 C_{\alpha\beta} \right) d x^\alpha d x^\beta,
   \label{metric}
\eea
where $a(\eta)$ is the cosmic scale factor and $\eta$ is the
conformal time defined as $c dt \equiv a d \eta$.
We set $c \equiv 1 \equiv \hbar$.
The Greek indices $\alpha$, $\beta$, $\gamma$, $\dots$ indicate
the space, and the Latic indices $a$, $b$, $c$, $\dots$ indicate
the spacetime.
The spacetime dependent variables $\alpha$, $\beta$, $\gamma$ and $\varphi$
are scalar-type perturbed order variables and $C_{\alpha\beta}$ is
a (transverse-tracefree) tensor-type variable.
Indices of $C_{\alpha\beta}$ are based on $g^{(3)}_{\alpha\beta}$
and a vertical bar
indicates a covariant derivative based on $g^{(3)}_{\alpha\beta}$.
The three-space metric $g^{(3)}_{\alpha\beta}$ indicates
the background comoving three-space part of the Robertson-Walker
metric which is spatially homogeneous and isotropic; some of its
specific representations are
\bea
   g^{(3)}_{\alpha\beta} d x^\alpha d x^\beta
   &=& {dr^2 \over 1 - K r^2}
       + r^2 \left( d \theta^2 + \sin^2{\theta} d \phi^2 \right)
   \nonumber \\
   &=& {1 \over \left( 1 + {K \over 4} \bar r^2 \right)^2}
       \left( d x^2 + d y^2 + d z^2 \right)
   \nonumber \\
   &=& d \bar \chi^2
       + \left( {1 \over \sqrt{K}} \sin{(\sqrt{K} \bar \chi)} \right)^2
       \left( d \theta^2 + \sin^2{\theta} d \phi^2 \right),
\eea
where we have
\bea
   & & r \equiv {\bar r \over 1 + {K \over 4} \bar r^2}, \quad
       \bar r \equiv \sqrt{x^2 + y^2 + z^2}, \quad
       \bar \chi \equiv \int^r {dr \over \sqrt{1-Kr^2}}.
\eea
The $K$ is the sign of the three-space curvature.

We ignore the vector-type perturbation (rotation) in this paper,
see \S \ref{sec:Discussions} for a summary.
Our metric convention follows Bardeen's \cite{Bardeen-1988}.
Our perturbation variables have some kinematic interpretations.
The kinematic quantites in the normal frame are \cite{Hwang-1991}:
\bea
   & & \theta = 3 H - \kappa, \quad
       \sigma_{\alpha\beta}
       = \chi_{,\alpha|\beta}
       - {1 \over 3} g_{\alpha\beta}^{(3)} \Delta \chi
       + a^2 \dot C^{(t)}_{\alpha\beta}, \quad
       a_\alpha = \alpha_{,\alpha}, \quad
       R^{(h)} = {1 \over a^2} \left[ 6 K - 4 \left( \Delta + 3 K \right)
       \varphi \right],
   \label{kinematic-quantities}
\eea
where we introduced
\bea
   & & \chi \equiv a \left( \beta + a \dot \gamma \right), \quad
       \kappa \equiv \delta K
       = 3 H \alpha - 3 \dot \varphi - {\Delta \over a^2} \chi.
   \label{chi-def}
\eea
An overdot denotes time derivative based on $t$, $H \equiv {\dot a \over a}$,
and $\Delta$ is a Laplacian operator based on $g^{(3)}_{\alpha\beta}$.
The $\theta$, $\sigma_{ab}$ and $a_a$ are the expansion scalar,
shear tensor and the acceleration vector, respectively;
in the normal-frame we have the vanishing rotation tensor
$\omega_{\alpha\beta} = 0$ based on the frame vector \cite{Covariant}.
{} From these we can interprete $\chi$, $\kappa$ and $\varphi$
as the perturbed shear, the perturbed expansion and the
perturbed curvature of the normal-frame vector.

As the energy-momentum tensor we consider an imperfect fluid form
including the scalar- and tensor-type perturbations
\bea
   & & T^0_0 = - \left( \bar \mu + \delta \mu \right), \quad
       T^0_\alpha = - \left( \mu + p \right) v_{,\alpha}, \quad
       T^\alpha_\beta = \left( \bar p + \delta p \right) \delta^\alpha_\beta
       + \Pi^\alpha_\beta,
   \label{Tab}
\eea
where $\Pi^\alpha_\beta$ is a tracefree anisotropic stress;
$\Pi^\alpha_\beta$ is based on $g^{(3)}_{\alpha\beta}$.
An overbar indicates the background order quantities;
we ignore it unless necessary.
The cosmological constant $\Lambda$ can be included by adding
$T^{(\Lambda)a}_{\;\;\;\;\;\; b} = - {\Lambda \over 8 \pi G} \delta^a_b$
to the energy-momentum tensor;
thus, $\Lambda$ can be included by adding
$\mu_\Lambda = - p_\Lambda = {\Lambda \over 8 \pi G}$
to the background fluid quantities $\mu$ and $p$.
The entropic perturbation $e$ is defined as
\bea
   & & e \equiv \delta p - c_s^2 \delta \mu, \quad
       c_s^2 \equiv \dot p / \dot \mu.
\eea
We decompose the anisotropic stress as
\bea
   & & \Pi_{\alpha\beta} \equiv {1 \over a^2} \left( \Pi_{,\alpha|\beta}
       - {1 \over 3} g^{(3)}_{\alpha\beta} \Delta \Pi \right)
       + \Pi_{\alpha\beta}^{(t)},
   \label{Pi}
\eea
where $\Pi_{\alpha\beta}^{(t)}$ is transverse and tracefree.
In an ideal fluid we have $e = 0$ and $\Pi = 0 = \Pi_{\alpha\beta}^{(t)}$.

\subsection{Gauge issue}

Here we summarize the behaviors of our perturbation variables under the
gauge transformation and our strategy of how to handle and use such
degrees of freedom as advantage.
Under the gauge transformation $\hat  x^a \equiv x^a + \tilde \xi^a (x^e)$
we have \cite{Bardeen-1988,Hwang-1991}
\bea
   & & \hat \alpha = \alpha - \dot \xi^t, \quad
       \hat \beta = \beta - {1 \over a} \xi^t
       + a \left( {\xi \over a} \right)^\cdot, \quad
       \hat \gamma = \gamma - {1 \over a} \xi, \quad
       \hat \varphi = \varphi - H \xi^t, \quad
       \hat \chi = \chi - \xi^t, \quad
       \hat \kappa = \kappa
       + \left( 3 \dot H + {\Delta \over a^2} \right) \xi^t,
   \nonumber \\
   & & \delta \hat \mu = \delta \mu - \dot \mu \xi^t, \quad
       \delta \hat p = \delta p - \dot p \xi^t, \quad
       \hat v = v - {1 \over a} \xi^t, \quad
       \hat \Pi = \Pi, \quad
       \delta \hat \phi = \delta \phi - \dot \phi \xi^t; \quad
       \hat C_{\alpha\beta} = C_{\alpha\beta}, \quad
       \hat \Pi_{\alpha\beta}^{({t})} = \Pi_{\alpha\beta}^{({t})},
   \label{GT}
\eea
where we used $\xi^0 \equiv {1 \over a} \xi^t$
and $\xi_\alpha \equiv \xi_{,\alpha}$;
$\bar \phi$ and $\delta \phi$ are the background and perturbed part of
a scalar field $\phi({\bf x}, t)$.
Thus, using $\chi$ instead of $\beta$ and $\gamma$ individually,
all our perturbation variables are spatially gauge-invariant.
However, all our scalar-type perturbation variables depend
on the temporal gauge transformation which will be used as
advantage in our gauge-ready strategy \cite{Hwang-1991}.
Temporal gauge fixing condition, fixing $\xi^t$,
applies only to the scalar-type perturbation.
To the linear-order, we can impose any one of the following temporal
gauge conditions to be valid at any spacetime point:
the synchronous gauge ($\alpha \equiv 0$),
the comoving gauge ($v \equiv 0$),
the zero-shear gauge ($\chi \equiv 0$),
the uniform-expansion gauge ($\kappa \equiv 0$),
the uniform-curvature gauge ($\varphi \equiv 0$),
the uniform-density gauge ($\delta \mu \equiv 0$),
the uniform-pressure gauge ($\delta p \equiv 0$),
the uniform-field gauge ($\delta \phi \equiv 0$).
Any linear combination of these gauge conditions which can give
a constraint on $\xi^t$ can be regarded as a suitable
temporal gauge condition, thus we have infinite number of
temporal gauge conditions available.
In the synchronous gauge we sets $\alpha \equiv 0$ in all coordinate
systems which leave nonvanishing $\xi^t ({\bf x})$ with general
dependence on spatial coordinate, see eq. (\ref{GT}).
Whereas in the other gauge conditions mentioned above we have
$\xi^t = 0$ after imposing any of the temporal gauge condition
in all coordinate systems, thus removing the gauge mode completely.
Thus, except for the synchronous gauge condition, each of the other
temporal gauge fixing conditions completely removes the temporal gauge mode.
Later, we will present our fundamental set of scalar-type perturbation
equations in a naturally spatially gauge-invariant form but
without fixing the temporal gauge conditions.
The equations will be arranged so that we can impose
any of our fundamental gauge conditions easily depending on the specific
problems we encounter; thus we suggestively call our approach
a gauge-ready formulation \cite{Hwang-1991}.

We introduce several gauge-invariant combinations:
\bea
   & & \varphi_\chi \equiv \varphi - H \chi, \quad
       \varphi_v \equiv \varphi - a H v, \quad
       \delta_v \equiv \delta - a {\dot \mu \over \mu} v, \quad
       \delta \phi_\varphi \equiv \delta \phi
       - {\dot \phi \over H} \varphi
       \equiv - {\dot \phi \over H} \varphi_{\delta \phi}, \quad
       v_\chi \equiv v - {1 \over a} \chi
       \equiv - {1 \over a} \chi_v,
   \label{GI-variables}
\eea
where $\delta \equiv \delta \mu / \mu$.
The gauge-invariant combination $\delta \phi_\varphi$
is equivalent to $\delta \phi$ in the uniform-curvature gauge which takes
$\varphi \equiv 0$ as the gauge condition, etc.
Using our notation for the gauge-invariant combinations
we can systematically construct and trace various
gauge-invariant combinations for a given variable, \cite{Hwang-1991}.
As in the last two examples in eq. (\ref{GI-variables}) our
notation of gauge-invariant variables allows algebra
connecting different expressions of the same gauge-invariant combinations.
Compared with the notation used by Bardeen in 1980 \cite{Bardeen-1980},
ignoring the harmonic functions, we have:
\bea
   & &
       \Phi_H \equiv \varphi_\chi, \quad
       \Phi_A \equiv \alpha_\chi, \quad
       \phi_m \equiv \varphi_v, \quad
       \epsilon_m \equiv \delta_v, \quad
       v^{(0)}_s \equiv k v_\chi, \quad
       p \pi_L^{(0)} \equiv \delta p, \quad
       p \pi_T^{(0)} \equiv - {\Delta \over a^2} \Pi.
   \label{Bardeen-notation}
\eea
Later we will use $\varphi_v$ (or $\varphi_{\delta \phi}$)
and $\varphi_\chi$ as the main variables.
Considering eqs. (\ref{kinematic-quantities},\ref{Tab}) we may regard
the gauge-invariant combinations $\varphi_v$ and $\varphi_\chi$ as
the perturbed three-space curvature ($\varphi$)
in the comoving gauge ($v \equiv 0$) and the zero-shear gauge ($\chi$),
respectively, based on the normal-frame vector field.

\subsection{Basic equations}
                                               \label{sec:Equations}

The background evolution is governed by
\bea
   & & H^2 = {8 \pi G \over 3} \mu - {K \over a^2} + {\Lambda \over 3}, \quad
       \dot H = - 4 \pi G \left( \mu + p \right) + {K \over a^2}.
   \label{BG-eqs}
\eea
To the perturbed order, the scalar-type perturbations are described by
 \cite{Bardeen-1988,Hwang-1991}
\bea
   & & \kappa \equiv 3 \left( - \dot \varphi + H \alpha \right)
       - {\Delta \over a^2} \chi,
   \label{G1} \\
   & & {\Delta + 3K \over a^2} \varphi + H \kappa = - 4 \pi G \delta \mu,
   \label{G2} \\
   & & \kappa + {\Delta + 3 K \over a^2} \chi
       = 12 \pi G a ( \mu + p ) v,
   \label{G3} \\
   & & \dot \chi + H \chi - \alpha - \varphi = 8 \pi G \Pi,
   \label{G4} \\
   & & \dot \kappa + 2 H \kappa
       + \left( 3 \dot H + {\Delta \over a^2} \right) \alpha
       = 4 \pi G \left( \delta \mu + 3 \delta p \right),
   \label{G5} \\
   & & \delta \dot \mu + 3 H \left( \delta \mu + \delta p \right)
       = \left( \mu + p \right) \left( \kappa
       - 3 H \alpha + {\Delta \over a} v \right),
   \label{G6} \\
   & & {\left[ a^4 ( \mu + p ) v \right]^\cdot \over a^4 ( \mu + p )}
       = {1 \over a} \left( \alpha + {\delta p \over \mu + p}
       + {2 \over 3} {\Delta + 3K \over a^2} {\Pi \over \mu + p} \right).
   \label{G7}
\eea
These follow from Einstein's equations and the energy and momentum
conservation equations;
eq. (\ref{G1}) is a definition of $\kappa$,
eqs. (\ref{G2}-\ref{G5}) follow from $G^0_0$, $G^0_\alpha$,
$G^\alpha_\beta - {1 \over 3} \delta^\alpha_\beta G^\gamma_\gamma$
and $G^\alpha_\alpha - G^0_0$ components of Einstein's equation,
respectively, and
eqs. (\ref{G6},\ref{G7}) follow from $T^b_{0;b} = 0$ and
$T^b_{\alpha;b} = 0$, respectively.
These equations are presented without fixing the temporal gauge
condition and using the spatially gauge-invariant variables only.
Thus, these are presented in a gauge-ready form which allows
us to choose the temporal gauge condition depending on the situation
as an advantage in handling the problem.
This set of equations was first presented by Bardeen in \cite{Bardeen-1988}.
As we are considering the most general imperfect fluid,
the above equations are valid even in the context of generalized
gravity theories we are considering:
the fluid quantities $\mu$, $p$, $\delta \mu$, $\delta p$, $(\mu + p)v$,
and $\Pi$ can be reinterpreted as the effective fluid quantities,
see below eq. (\ref{GFE}) and \cite{Hwang-GGT-CQG-1990,Hwang-1991}.

In the case of the gravitational wave we have
\bea
   & & \ddot C^\alpha_\beta + 3 H \dot C^\alpha_\beta
       - {\Delta - 2 K \over a^2} C^\alpha_\beta
       = 8 \pi G \Pi^{(t)\alpha}_{\;\;\;\; \beta}.
   \label{GW-eq}
\eea

In eqs. (\ref{BG-eqs}-\ref{G7},\ref{GW-eq}) we presented
the complete sets of background
and the perturbed equations for a general imperfect fluid.
In the case of fluid we include the most general type of imperfect-fluid
contributions for both the background and perturbation.
In such forms the equations are generally valid
for the case of a scalar field and the other generalized gravity
theories, by re-interpreting the fluid quantities as the effective
fluid quantities.
That is, by arranging the gravitational field equation in the following form
\bea
   & & G^a_b = 8 \pi G T^a_b
   \label{GFE}
\eea
we re-interprete $T^a_b$ as the effective
energy-momentum tensor \cite{Hwang-GGT-CQG-1990}.
Thus, in those generalized theories we will present only
the effective fluid quantities which together with the general equations
derived in the fluid in eqs. (\ref{BG-eqs}-\ref{G7})
provide complete set of equations.

\section{Equations in two gauges}
                                               \label{sec:two-gauges}

\subsection{Fluid}
                                               \label{sec:Fluid}

We introduce the Field-Shepley combination \cite{Field-Shepley-1968}
\bea
   & & \Phi \equiv \varphi_v - {K /a^2 \over 4 \pi G ( \mu + p) } \varphi_\chi.
   \label{Phi-def}
\eea
We can derive \cite{H-Hydro-1999}
\bea
   & & \dot \Phi = {H \over 4 \pi G ( \mu + p) } {c_s^2 \Delta \over a^2}
       \varphi_\chi
       - {H \over \mu + p} \left( e
       + {2 \over 3} {\Delta \over a^2} \Pi \right),
   \label{dot-Phi-eq-Fluid} \\
   & & {H \over a} \left( {a \over H} \varphi_\chi \right)^\cdot
       = {4 \pi G ( \mu + p) \over H} \Phi - 8 \pi G H \Pi.
   \label{dot-Psi-eq-Fluid}
\eea
Equation (\ref{dot-Phi-eq-Fluid}) follows from taking a time derivative
of eq. (\ref{Phi-def}) and using eqs. (\ref{G1}-\ref{G4},\ref{G7});
we also need $v_\chi = - {1 \over a} \chi_v$ in eq. (\ref{GI-variables}).
Equation (\ref{dot-Psi-eq-Fluid}) follows from eq. (\ref{G1}) and using
eqs. (\ref{G3},\ref{G4}) with
$v_\chi = v - {1 \over a} \chi = - {1 \over aH} (\varphi_v - \varphi_\chi)$
which follows from eq. (\ref{GI-variables}).
We can combine eqs. (\ref{dot-Phi-eq-Fluid},\ref{dot-Psi-eq-Fluid})
to make closed form second-order differential equations for
both $\Phi$ and $\varphi_\chi$
\bea
   & & {H^2 c_s^2 \over a^3 (\mu + p)}
       \left\{ {a^3 (\mu + p) \over H^2 c_s^2}
       \left[ \dot \Phi
       + {H \over \mu + p} \left( e
       + {2 \over 3} {\Delta \over a^2} \Pi \right) \right]
       \right\}^\cdot
       = c_s^2 {\Delta \over a^2} \left( \Phi
       - 2 {H^2 \over \mu + p} \Pi \right),
   \label{ddot-Phi-eq-Fluid} \\
   & & {\mu + p \over H} \left[
       {H^2 \over a (\mu + p)} \left( {a \over H} \varphi_\chi \right)^\cdot
       + 8 \pi G {H^2 \over \mu + p} \Pi \right]^\cdot
       = c_s^2 {\Delta \over a^2} \varphi_\chi
       - 4 \pi G \left( e + {2 \over 3} {\Delta \over a^2} \Pi \right).
   \label{ddot-Psi-eq-Fluid}
\eea
In an ideal fluid we have $e = 0 = \Pi$.
{}For a pressureless medium we have $c_s^2 = 0$,
thus instead of eqs. (\ref{dot-Phi-eq-Fluid},\ref{ddot-Phi-eq-Fluid}) we have
\bea
   & & \dot \Phi = 0.
   \label{ddot-Phi-eq-pressureless}
\eea
Equations (\ref{dot-Phi-eq-Fluid}-\ref{ddot-Psi-eq-Fluid}) show basic
forms of equations valid even in the scalar field and generalized gravity
theories to be considered in this paper.
Unified forms will be presented in \S \ref{sec:CE}.

In the fluid context it is often convenient to have equation for
density perturbation.
The most convenient (i.e., similar to the Newtonian) form is available
in the comoving gauge,
thus equivalently using a gauge-invariant combination $\delta_v$,
\cite{HN-Newtonian-1999}.
Using the Poisson's relation
\bea
   & & - {\Delta + 3 K \over a^2} \varphi_\chi = 4 \pi G \mu \delta_v,
   \label{Poisson-eq}
\eea
which follows from eqs. (\ref{G2},\ref{G3}), eq. (\ref{ddot-Psi-eq-Fluid})
gives \cite{HN-Newtonian-1999}
\bea
   & & {\mu + p \over a^2 H \mu} \left[
       {H^2 \over a (\mu + p)} \left( {a^3 \mu \over H} \delta_v \right)^\cdot
       - 2 {H^2 \over \mu + p} \left( \Delta + 3K \right) \Pi \right]^\cdot
       = c_s^2 {\Delta \over a^2} \delta_v
       + {\Delta + 3K \over a^2} {1 \over \mu}
       \left( e + {2 \over 3} {\Delta \over a^2} \Pi \right).
   \label{ddot-delta_v-eq}
\eea

{}For the tensor mode eq. (\ref{GW-eq}) gives
\bea
   & & {1 \over a^3} \left( a^3 \dot C_{\alpha\beta} \right)^\cdot
       - {\Delta - 2 K \over a^2} C_{\alpha\beta}
       = 8 \pi G \Pi_{\alpha\beta}^{(t)}.
   \label{GW-eq-Fluid}
\eea
The fluid perturbation and the gravitational wave were
studied in the context of the synchronous gauge ($\alpha \equiv 0$)
by Lifshitz \cite{Lifshitz-1946}.
The zero-shear gauge ($\chi \equiv 0$) to handle the gravitational potential
($\varphi$) and the comoving gauge ($v \equiv 0$) for the density
perturbation ($\delta$), were first studied by
Harrison \cite{Harrison-1967} and Nariai \cite{Nariai-1969}, respectively.

\subsection{Field}
                                               \label{sec:MSF}

We consider an action for a minimally coupled scalar field
\cite{Mukhanov-1985,Mukhanov-1988,Hwang-MSF}
\bea
   & & S = \int d^4 x \sqrt{-g} \left[ {1 \over 16 \pi G} R
       - {1 \over 2} \phi^{,c} \phi_{,c} - V (\phi) \right].
   \label{MSF-action}
\eea
The gravitational field equation and the equation of motion are
\bea
   & & G_{ab} = 8 \pi G \left(
       \phi_{,a} \phi_{,b}
       - {1 \over 2} \phi^{,c} \phi_{,c} g_{ab}
       - V g_{ab} \right),
   \label{GFE-MSF} \\
   & & \Box \phi - V_{,\phi} = 0.
   \label{EOM-MSF}
\eea
Equations (\ref{BG-eqs}-\ref{G7}) remain valid with the following
background and perturbed order fluid quantities
\bea
   & & \mu \equiv {1 \over 2} \dot \phi^2 + V, \quad
       p \equiv {1 \over 2} \dot \phi^2 - V,
   \label{MSF-fluid-BG} \\
   & & \delta \mu \equiv \dot \phi \delta \dot \phi
       - \dot \phi^2 \alpha + V_{,\phi} \delta \phi, \quad
       \delta p \equiv \dot \phi \delta \dot \phi
       - \dot \phi^2 \alpha - V_{,\phi} \delta \phi, \quad
       v \equiv {1 \over a} { \delta \phi \over \dot \phi },
       \quad
       \Pi = 0 = \Pi^{(t)\alpha}_{\;\;\;\; \beta},
   \label{MSF-fluid-pert}
\eea
where we expanded
$\phi ({\bf x}, t) = \bar \phi (t) + \delta \phi ({\bf x}, t)$.
Additionally, from eq. (\ref{EOM-MSF}) we have the background
and perturbed equation of motion
\bea
   & & \ddot \phi + 3 H \dot \phi + V_{,\phi} = 0,
   \\
   & & \delta \ddot \phi + 3 H \delta \dot \phi
       - {\Delta \over a^2} \delta \phi + V_{,\phi\phi} \delta \phi
       = \dot \phi \left( \kappa + \dot \alpha \right)
       + \left( 2 \ddot \phi + 3 H \dot \phi \right) \alpha.
   \label{EOM-MSF-pert}
\eea

As $\delta \phi = 0$ implies $v = 0$, the uniform-field gauge
($\delta \phi \equiv 0$) is equivalent to the comoving gauge ($v \equiv 0$),
thus $\varphi_v = \varphi_{\delta \phi}$.
Thus, eq. (\ref{Phi-def}) becomes
\bea
   & & \Phi \equiv \varphi_{\delta \phi}
       - {K/a^2 \over 4 \pi G \dot \phi^2} \varphi_\chi.
   \label{Phi-def-MSF}
\eea
We have
\bea
   & & \dot \Phi = {H \over 4 \pi G \dot \phi^2 } {c_A^2 \Delta \over a^2}
       \varphi_\chi,
   \label{dot-Phi-eq-MSF} \\
   & & {H \over a} \left( {a \over H} \varphi_\chi \right)^\cdot
       = {4 \pi G \dot \phi^2 \over H} \Phi,
   \label{dot-Psi-eq-MSF}
\eea
where
\bea
   & & c_A^2 \equiv 1 + 3 (1 - c_s^2) K \Delta^{-1}, \quad
       c_s^2 \equiv {\dot p \over \dot \mu}
             = -1 - {2 \ddot \phi \over 3 H \dot \phi}.
\eea
Equation (\ref{dot-Phi-eq-MSF}) follows from taking a time derivative of
eq. (\ref{Phi-def-MSF}) and using eqs. (\ref{G1}-\ref{G4});
we also need $\delta \mu_{\delta\phi} = - \dot \phi^2 \alpha_{\delta \phi}$
which follows from eq. (\ref{MSF-fluid-pert}).
Equation (\ref{dot-Psi-eq-MSF}) follows from eq. (\ref{G1}) and
using eqs. (\ref{G3},\ref{G4}) with
$\delta \phi_\chi = \delta \phi - \dot \phi \chi
 = {\dot \phi \over H} ( - \varphi_{\delta \phi} + \varphi_\chi )$
which follows from eq. (\ref{GI-variables}).
Equations (\ref{dot-Phi-eq-MSF},\ref{dot-Psi-eq-MSF}) also follow from
eqs. (\ref{dot-Phi-eq-Fluid},\ref{dot-Psi-eq-Fluid}) using
the effective fluid quantities in eqs.
(\ref{MSF-fluid-BG},\ref{MSF-fluid-pert}):
from eqs. (\ref{MSF-fluid-BG},\ref{MSF-fluid-pert}) we have
\bea
   & & e = \left( 1 - c_s^2 \right) \delta \mu_{\delta \phi}
       = - \left( 1 - c_s^2 \right) {\Delta + 3 K \over 4 \pi G a^2}
       \varphi_\chi,
   \label{e-MSF}
\eea
where we used eq. (\ref{Poisson-eq}), and
$\delta \mu_{\delta \phi} = \delta \mu_v$.
Using eq. (\ref{e-MSF}) and $\Pi = 0$ in
eqs. (\ref{dot-Phi-eq-Fluid},\ref{dot-Psi-eq-Fluid})
we can derive eqs. (\ref{dot-Phi-eq-MSF},\ref{dot-Psi-eq-MSF}).
By combining eqs. (\ref{dot-Phi-eq-MSF},\ref{dot-Psi-eq-MSF}) we have
\bea
   & & {H^2 c_A^2 \over a^3 \dot \phi^2}
       \left[ {a^3 \dot \phi^2 \over H^2 c_A^2} \dot \Phi \right]^\cdot
       = c_A^2 {\Delta \over a^2} \Phi,
   \label{ddot-Phi-eq-MSF} \\
   & & {\dot \phi^2 \over H} \left[ {H^2 \over a \dot \phi^2}
       \left( {a \over H} \varphi_\chi \right)^\cdot \right]^\cdot
       = c_A^2 {\Delta \over a^2} \varphi_\chi,
   \label{ddot-Psi-eq-MSF}
\eea
which can be compared with the ideal fluid (thus sets $e \equiv 0 \equiv \Pi$)
equations in eqs. (\ref{ddot-Phi-eq-Fluid},\ref{ddot-Psi-eq-Fluid}).
Thus, we have an interesting conclusion:
\begin{quote}
``{\sl Compared with an ideal fluid, the minimally coupled scalar field
       effectively has $c_s^2$ replaced by $c_A^2$ which
       becomes $1$ for $K = 0$.}''
\end{quote}
The $c_A$ has the role of wave speed of the perturbed field
and the simultaneously excited metric;
interpretation of $c_A$ as the wave speed is properly valid only for $K = 0$.
Thus, for $K = 0$ the wave propagation speed becomes $1$.
In the minimally coupled scalar field $c_s$ cannot be interpreted
as the wave propagation speed.
This is because the scalar field has a non-vanishing  entropic perturbation
as in eq. (\ref{e-MSF}), thus cannot be interpreted as an ideal fluid.

{}For the tensor mode, eq. (\ref{GW-eq-Fluid}) remains
valid in the field situation with $\Pi_{\alpha\beta}^{(t)} = 0$.
That is, the presence of minimally coupled scalar field (or fields)
does not directly affect the equation of tensor-type perturbation.

\subsection{Generalized $f(\phi,R)$ gravity}
                                               \label{sec:GGT}

We consider an action
\cite{Hwang-GGT-CQG-1990,Hwang-1991,HN-CMBR-2002,H-GGT-1995,HN-GGT-1996}
\bea
   & & S = \int d^4 x \sqrt{-g} \left[ {1 \over 2} f(\phi, R)
       - {1 \over 2} \omega (\phi) \phi^{,c} \phi_{,c} - V (\phi)
       + L_{(m)} + L_{(c)} \right],
   \label{GGT-action}
\eea
where $L_{(c)}$ represents additional correction terms to be considered in
\S \ref{sec:String-corrections} and \ref{sec:String-axion}.
This action without $L_{(c)}$ includes the following gravity theories as subset:
$f(R)$ gravity which includes $R^2$ gravity,
the scalar-tensor theory which includes the Jordan-Brans-Dicke theory,
the non-minimally coupled scalar field,
the induced gravity,
the low-energy effective action of string theory, etc, see \cite{HN-GGT-1996}.
Although this generalized action by itself does not have much physical meaning,
there are some advantages by analysing perturbations in this general context:
for example, our results will be valid considering {\it transitions}
from one type of gravity theory to the other.
The gravitational field equation and the equation of motion are
\bea
   & & G_{ab} = {1 \over F} \left[
       T^{(m)}_{ab}
       + \omega \left( \phi_{,a} \phi_{,b}
       - {1 \over 2} \phi^{,c} \phi_{,c} g_{ab} \right)
       + {1 \over 2} \left( f - RF - 2 V \right) g_{ab}
       + F_{,a;b} - \Box F g_{ab}
       + T^{(c)}_{ab}
       \right],
   \label{GFE-GGT} \\
   & & \Box \phi
       + {1 \over 2 \omega} \left(
       \omega_{,\phi} \phi^{,c} \phi_{,c}
       + f_{,\phi} - 2 V_{,\phi} \right)
       = {1 \over 2 \omega} T^{(c)},
   \label{EOM-GGT}
\eea
where $F \equiv {\partial f \over \partial R}$.
We have
\bea
   & & \delta \left( \sqrt{-g} L_{(m)} \right)
       \equiv {1 \over 2} \sqrt{-g} T_{(m)}^{ab} \delta g_{ab}, \quad
       \delta \left( \sqrt{-g} L_{(c)} \right)
       \equiv {1 \over 2} \sqrt{-g} T_{(c)}^{ab} \delta g_{ab}.
\eea

Equations (\ref{BG-eqs}-\ref{GW-eq}) remain valid with
the following effective fluid quantities
\bea
   8 \pi G \mu
   &=& {1 \over F} \left( {1 \over 2} \omega \dot \phi^2
       + {R F - f + 2 V \over 2} - 3 H \dot F
       - T^{(c)0}_{\;\;\;\;\,0} \right),
   \nonumber \\
   8 \pi G p
   &=& {1 \over F} \left( {1 \over 2} \omega \dot \phi^2
       - {R F - f + 2 V \over 2} + \ddot F + 2 H \dot F
       + {1 \over 3} T^{(c)\alpha}_{\;\;\;\;\,\alpha} \right),
   \label{BG-fluid-GGT} \\
   8 \pi G \delta \mu
   &=& {1 \over F} \Bigg[ \omega \dot \phi \delta \dot \phi
       + {1 \over 2} \left( \omega_{,\phi} \dot \phi^2
       - f_{,\phi} + 2 V_{,\phi} \right) \delta \phi - 3 H \delta \dot F
       + \left( 3 \dot H + 3 H^2 + {\Delta \over a^2} \right) \delta F
   \nonumber \\
   & &
       + \left( 3 H \dot F - \omega \dot \phi^2 \right) \alpha
       + \dot F \kappa
       - \left( \delta T^{(c)0}_{\;\;\;\;\,0}
       - {\delta F \over F} T^{(c)0}_{\;\;\;\;\,0} \right)
       \Bigg],
   \nonumber \\
   8 \pi G \delta p
   &=& {1 \over F} \Bigg[ \omega \dot \phi \delta \dot \phi
       + {1 \over 2} \left( \omega_{,\phi} \dot \phi^2 + f_{,\phi}
       - 2 V_{,\phi} \right) \delta \phi
       + \delta \ddot F + 2 H \delta \dot F
       + \left( - \dot H - 3 H^2
       - {2 \over 3} {\Delta + 3 K \over a^2} \right) \delta F
   \nonumber \\
   & &
       - \dot F \dot \alpha
       - \left( \omega \dot \phi^2 + 2 \ddot F + 2 H \dot F \right) \alpha
       - {2 \over 3} \dot F \kappa
       + {1 \over 3} \left( \delta T^{(c)\alpha}_{\;\;\;\;\,\alpha}
       - {\delta F \over F} T^{(c)\alpha}_{\;\;\;\;\,\alpha} \right)
       \Bigg],
   \nonumber \\
   8 \pi G T^0_\alpha
   &=& {1 \over F} \left[
       {1 \over a} \left( - \omega \dot \phi \delta \phi
       - \delta \dot F + H \delta F + \dot F \alpha \right)_{,\alpha}
       + T^{(c)0}_{\;\;\;\;\; \alpha} \right],
   \nonumber \\
   8 \pi G \Pi^\alpha_\beta
   &=& {1 \over F} \left[ {1 \over a^2} \left( \nabla^\alpha \nabla_\beta
       - {1 \over 3} \delta^\alpha_\beta \Delta \right)
       \left( \delta F - \dot F \chi \right)
       - \dot F \dot C^\alpha_\beta
       + \delta T^{(c)\alpha}_{\;\;\;\;\,\beta}
       - {1 \over 3} \delta^\alpha_\beta \delta T^{(c)\gamma}_{\;\;\;\;\,\gamma}
       \right],
   \label{pert-fluid-GGT}
\eea
where we have
\bea
   & & R = 6 \left( 2 H^2 + \dot H + {K \over a^2} \right),
   \label{R} \\
   & & \delta R = 2 \left[ - \dot \kappa - 4 H \kappa
       - \left( {\Delta \over a^2} + 3 \dot H \right) \alpha
       - 2 {\Delta + 3K \over a^2} \varphi \right]
       = \delta \mu - 3 \delta p.
   \label{delta-R}
\eea
We have set $T_{ab}^{(m)} = 0$.
The scalar- and tensor-type fluid quantities can be read from
eqs. (\ref{Tab},\ref{Pi}) as
\bea
   & & \left( \mu + p \right) v = - \Delta^{-1} \nabla^\alpha T^0_\alpha,
   \nonumber \\
   & & \Pi = {3 \over 2} a^2 \Delta^{-1} \left( \Delta + 3K \right)^{-1}
       \left( \Pi^{\alpha|\beta}_{\beta\;\;\;\alpha} \right), \quad
       \Pi^{(t)\alpha}_{\;\;\;\;\;\beta}
       = \Pi^\alpha_\beta
       - {1 \over a^2} \left( \nabla^\alpha \nabla_\beta
       - {1 \over 3} \delta^\alpha_\beta \Delta \right) \Pi, \quad
       \Pi^\alpha_\beta = T^\alpha_\beta
       - {1 \over 3} \delta^\alpha_\beta T^\gamma_\gamma.
   \label{pert-fluid-decomposition}
\eea
The comoving gauge ($v \equiv 0$) differs from the uniform-field gauge
($\delta \phi \equiv 0$), thus $\varphi_{\delta \phi} \neq \varphi_v$.
The equation of motion gives
\bea
   & & \ddot \phi + 3 H \dot \phi + {1 \over 2 \omega}
       \left( \omega_{,\phi} \dot \phi^2 - f_{,\phi} + 2 V_{,\phi} \right)
       = - {1 \over 2 \omega} T^{(c)}.
   \label{BG-EOM-GGT} \\
   & & \delta \ddot \phi + \left( 3 H
       + {\omega_{,\phi} \over \omega} \dot \phi \right) \delta \dot \phi
       + \left[ - {\Delta \over a^2}
       + \left( {\omega_{,\phi} \over \omega} \right)_{,\phi}
       {\dot \phi^2 \over 2}
       + \left( {-f_{,\phi} + 2 V_{,\phi} \over 2 \omega} \right)_{,\phi}
       \right] \delta \phi
   \nonumber \\
   & & \qquad
       = \dot \phi \dot \alpha + \left( 2 \ddot \phi + 3 H \dot \phi
       + { \omega_{,\phi} \over \omega} \dot \phi^2 \right) \alpha
       + \dot \phi \kappa + {1\over 2 \omega} F_{,\phi} \delta R
       - {1 \over 2 \omega} \left( \delta T^{(c)}
       - {\omega_{,\phi} \over \omega} \delta \phi T^{(c)} \right).
   \label{pert-EOM-GGT}
\eea
We have additionally located $T^{(c)}_{ab}$ and $T^{(c)}$
terms for later consideration of the string theory correction terms
in \S \ref{sec:String-corrections} and \ref{sec:String-axion}.
In this subsection we ignore these correction terms.
{}From eqs. (\ref{pert-fluid-GGT},\ref{pert-fluid-decomposition},\ref{Tab})
we have
\bea
   & & 8 \pi G ( \mu + p ) v = {1 \over aF} \left( \omega \dot \phi \delta \phi
       + \delta \dot F - H \delta F - \dot F \alpha \right),
   \nonumber \\
   & & 8 \pi G \Pi = { 1\over F} \left( \delta F - \dot F \chi \right), \quad
       8 \pi G \Pi^{(t)\alpha}_{\;\;\;\;\;\beta}
       = - {\dot F \over F} C^{\alpha}_{\beta}.
   \label{pert-fluid-GGT2}
\eea

Instead of eq. (\ref{Phi-def-MSF}) we introduce a more generalized form
\bea
   & & \Phi \equiv \varphi_{\delta \phi} - 2 {K \over a^2}
       {F \over \omega \dot \phi^2 + {3 \dot F^2 \over 2 F}} \Psi, \quad
       \Psi \equiv \varphi_\chi + {\delta F_\chi \over 2 F}.
   \label{Psi-def}
\eea
{}From eqs. (\ref{G1}-\ref{G3}) and eqs. (\ref{G1},\ref{G3},\ref{G4}),
respectively, we can derive
\bea
   & & \dot \Phi
       = {2 H F + \dot F \over \omega \dot \phi^2 + {3 \dot F^2 \over 2 F}}
       {c_A^2 \Delta \over a^2} \Psi,
   \label{dot-Phi-eq-GGT} \\
   & & {H + {\dot F \over 2F} \over aF}
       \left( {a F \over H + {\dot F \over 2F}} \Psi \right)^\cdot
       = {\omega \dot \phi^2 + {3 \dot F^2 \over 2F} \over 2 H F + \dot F}
       \Phi,
   \label{dot-Psi-eq-GGT}
\eea
where
\bea
   & & c_A^2 \equiv 1 + \left( 6 + {2 {\ddot \phi \over \dot \phi}
       - 3 {\dot F \over F} + {\dot E \over E} \over
       H + {\dot F \over 2 F} } \right) K \Delta^{-1}, \quad
       E \equiv F \left( \omega + {3 \dot F^2 \over 2 F \dot \phi^2} \right).
  \label{E-def}
\eea
In order to derive eqs. (\ref{dot-Phi-eq-GGT},\ref{dot-Psi-eq-GGT})
we follow the same algebraic procedure as needed in the scalar field.
By combining eqs. (\ref{dot-Phi-eq-GGT},\ref{dot-Psi-eq-GGT}) we have
\bea
   & & {\left( H + {\dot F \over 2F} \right)^2 c_A^2
       \over a^3 \left( \omega \dot \phi^2 + {3 \dot F^2 \over 2F} \right)}
       \left[ {a^3 \left( \omega \dot \phi^2 + {3 \dot F^2 \over 2F} \right)
       \over \left( H + {\dot F \over 2F} \right)^2 c_A^2} \dot \Phi
       \right]^\cdot
       = c_A^2 {\Delta \over a^2} \Phi,
   \label{ddot-Phi-eq-GGT} \\
   & & {\omega \dot \phi^2 + {3 \dot F^2 \over 2F} \over
       H F + {1 \over 2} \dot F} \left[
       { \left( H + {\dot F \over 2F} \right)^2 \over
       a \left( \omega \dot \phi^2 + {3 \dot F^2 \over 2F} \right) }
       \left( {a F \over H + {\dot F \over 2F}} \Psi \right)^\cdot \right]^\cdot
       = c_A^2 {\Delta \over a^2} \Psi.
   \label{ddot-Psi-eq-GGT}
\eea
We notice that $c_A$ can be interpreted as a wave speed of
the perturbed field as well as the simultaneously excited metric;
for $K = 0$ we have
\bea
   & & c_A = 1,
\eea
and only in this case we can properly interprete $c_A$ as the wave speed;
$c_A^2$ clearly differs from $c_s^2 \equiv \dot p/\dot \mu$.

This unified result is valid for the gravities in the forms
either (i) $F = F (\phi)$ or (ii) $F = F(R)$.
In the case of $f = f(R)$ without the field the results in the above
remain valid with the following prescription:
we remove $\phi$ ($\omega$ as well) thus we have $E = {3 \over 2} \dot F^2$,
and set $\varphi_{\delta \phi} = \varphi_{\delta F}$.
In the general case with $F = F (\phi, R)$ the situation corresponds
to the two component medium, see \S \ref{sec:CT}.

Using
\bea
   & & \delta F_\chi = {\dot F \over H}
       \left( \varphi_\chi - \varphi_{\delta \phi} \right),
   \label{delta-F_chi}
\eea
we have
\bea
   & & \varphi_\chi
       = {H \Psi + {\dot F \over 2 F} \varphi_{\delta \phi}
       \over H + {\dot F \over 2 F}}.
   \label{varphi_chi}
\eea
As we are considering the linear theory, if the solution of one variable
is known in any one gauge, solutions of all the other variables
in the same gauge as well as in any other gauge can be derived easily
through linear algebra; our gauge-ready form equations in
eqs. (\ref{G1}-\ref{G7}) and our convention of the gauge invariant variables
in eq. (\ref{GI-variables}) are useful to
derive the remaining variables systematically.

{}For the tensor mode we have
\bea
   & & {1 \over a^3 F} \left( a^3 F \dot C_{\alpha\beta} \right)^\cdot
       - {\Delta - 2 K \over a^2} C_{\alpha\beta} = 0.
   \label{GW-eq-GGT}
\eea
This equation is valid for general algebraic function of $f(\phi, R)$.
In the context of fluid formulation in eq. (\ref{GW-eq-Fluid}), the
$F$ term in eq. (\ref{GW-eq-GGT}) comes from the nonvanishing
effective anisotropic stress of our generalized gravity theory
in eq. (\ref{pert-fluid-GGT2}).

\subsection{Conformal transformation}
                                                   \label{sec:CT}

Although direct derivation of eqs. (\ref{dot-Phi-eq-GGT},\ref{dot-Psi-eq-GGT})
is not complicated there is another simple way.
The gravity theory in eq. (\ref{GGT-action}) can be transformed to
Einstein's gravity through a conformal rescaling of the metric
and rescaling of the field.
The end result is Einstein's gravity with only complications
appearing in the modified form of the field potential.
Thus, using such transformation properties we can derive
the equations in generalized gravity from equations in \S \ref{sec:MSF}.

We have studied the conformal transformation properties in
\cite{Hwang-GGT-CQG-1990,Hwang-CT-1997},
and in most general form in the Appendix A of \cite{HN-CMBR-2002}.
Under the conformal transformation
\bea
   & & \hat g_{ab} = \Omega^2 g_{ab}, \quad
       \Omega \equiv \sqrt{8 \pi G F} \equiv e^{\psi/\sqrt{6}},
\eea
with
\bea
   & & d \hat \phi \equiv \sqrt{{1 \over 8 \pi G}
       \left( {\omega \over F} d \phi^2 + d \psi^2 \right)},
\eea
eq. (\ref{GGT-action}) transforms to Einstein's gravity
with a modified potential
\bea
   & & \hat V = {1 \over (16 \pi G F)^2} \left( 2 V - f + R F \right).
\eea
The form of $d \hat \phi$ implies that, in order to have
Einstein's gravity with one minimally coupled scalar field
we need certain condition on the form of $f(\phi, R)$.
We need either $f = F(\phi) R$ or a pure $f(R)$ gravity without the field;
if we have $f$ a nonlinear function of $R$ with a field involved, we
will have Einstein's gravity with two scalar fields.

Under such a transformation, with
$\Omega \equiv \bar \Omega (1 + \delta \Omega)$,
we have, to the background order
\bea
   & & \hat a = a \Omega, \quad
       d \hat t = \Omega dt, \quad
       \hat H = {1 \over \Omega} \left( H + {\dot \Omega \over \Omega} \right),
       \quad
       \dot {\hat \phi} = \sqrt{ {1 \over 8 \pi G} \left( {\omega \over F}
       \dot \phi^2 + {3 \dot F^2 \over 2 F^2} \right) },
   \label{CT-BG}
\eea
and to the perturbed order
\bea
   & & \hat \varphi = \varphi + \delta \Omega, \quad
       \hat \alpha = \alpha + \delta \Omega, \quad
       \hat \chi = \Omega \chi, \quad
       {\delta \hat \phi \over \dot {\hat \phi}}
       = {\delta \phi \over \dot \phi}
       = {\delta F \over \dot F},
   \label{CT-pert}
\eea
where
\bea
   & & \Omega = \sqrt{ 8 \pi G F }, \quad
       \delta \Omega = {\delta F \over 2 F}.
\eea
The following variables are invariant under the conformal transformation
\bea
   & & d \eta, \quad
       \Delta, \quad
       k^2, \quad
       \varphi_{\delta \phi} = - {H \over \dot \phi} \delta \phi_\varphi, \quad
       C_{\alpha\beta},
   \label{CT-inv}
\eea
where $k$ is the comoving wave number with $\Delta = - k^2$
in the Fourier space.
It would be a trivial exercise to derive
eqs. (\ref{dot-Phi-eq-GGT},\ref{dot-Psi-eq-GGT}) from
eqs. (\ref{dot-Phi-eq-MSF},\ref{dot-Psi-eq-MSF}) using
eqs. (\ref{CT-BG}-\ref{CT-inv}).
We can also show that under the conformal transformation we have
$\hat \Phi = \Phi$ and $\hat \varphi_{\hat \chi} = \Psi$.
Notice that $\varphi_{\delta \phi}$ and $C_{\alpha\beta}$ are
invariant under the conformal transformation.
{}From this result we can draw an {\it important conclusion} that
\begin{quote}
``{\sl the power spectra (both amplitudes and spectral slopes)
       based on $\varphi_{\delta \phi}$ and $C_{\alpha\beta}$ derived
       in the context of any given frame is valid in any other frame.}''
\end{quote}

However, for the two other types of gravity theories to be considered
in the following subsections we have no such transformation property available.
We have to derive the equations directly from the gravity theories
and equations of motion.
The same algebraic procedures needed for the scalar field also apply
to these gravity theories in a rather exact manner.

\subsection{Tachyonic generalization}
                                               \label{sec:Tachyon}

We consider an action \cite{HN-Tachyon-2002}
\bea
   & & S = \int d^4 x \sqrt{-g} \left[ {1 \over 2} f (R, \phi, X)
       + L_{(m)} + L_{(c)} \right],
   \label{Tachyon-action}
\eea
where $X \equiv {1 \over 2} \phi^{,c} \phi_{,c}$,
and $L_{(c)}$ includes additional correction terms to be considered in
\S \ref{sec:String-corrections} and \ref{sec:String-axion}.
This action includes the generalized $f(\phi, R)$ gravity
in eq. (\ref{GGT-action}) as a case.
The gravitational field equation and the equation of motion are
\bea
   & & G_{ab} = {1 \over F} \left[
       T^{(m)}_{ab}
       + {1 \over 2} \left( f - R F \right) g_{ab}
       + F_{,a;b}
       - \Box F g_{ab}
       - {1 \over 2} f_{,X} \phi_{,a} \phi_{,b}
       + T^{(c)}_{ab}
       \right],
   \label{GFE-Tachyon} \\
   & & \left( f_{,X} \phi^{,c} \right)_{;c} = f_{,\phi} - T^{(c)}.
   \label{EOM-Tachyon}
\eea

Equations (\ref{BG-eqs}-\ref{GW-eq}) remain valid with
the following effective fluid quantities
\bea
   8 \pi G \mu
   &=& {1 \over F} \left( f_{,X} X + {F R - f \over 2}
       - 3 H \dot F
       - T^{(c)0}_{\;\;\;\;\,0}
       \right),
   \nonumber \\
   8 \pi G p
   &=& {1 \over F} \left(
       - {F R -f \over 2} + \ddot F + 2 H \dot F
       + {1 \over 3} T^{(c)\alpha}_{\;\;\;\;\,\alpha}
       \right),
   \label{BG-fluid-tachyon} \\
   8 \pi G \delta \mu
   &=& {1 \over F} \Bigg[
       - {1 \over 2} \left( f_{,\phi} \delta \phi + f_{,X} \delta X \right)
       - {1 \over 2} \dot \phi^2 \left( F_{,X} \delta R
       + f_{,X \phi} \delta \phi + f_{,XX} \delta X \right)
   \nonumber \\
   & &
       - f_{,X} \dot \phi \delta \dot \phi
       - 3 H \delta \dot F
       + \left( 3 \dot H + 3 H^2  + {\Delta \over a^2} \right) \delta F
       + \dot F \kappa
       + \left( 3 H \dot F + f_{,X} \dot \phi^2 \right) \alpha
       - \left( \delta T^{(c)0}_{\;\;\;\;\,0}
       - {\delta F \over F} T^{(c)0}_{\;\;\;\;\,0} \right)
       \Bigg],
   \nonumber \\
   8 \pi G \delta p
   &=& {1 \over F} \Bigg[
       {1 \over 2} \left( f_{,\phi} \delta \phi + f_{,X} \delta X \right)
       + \delta \ddot F + 2 H \delta \dot F
       + \left( - \dot H - 3 H^2 - {2 \over 3} {\Delta + 3 K \over a^2}
       \right) \delta F
   \nonumber \\
   & &
       - {2 \over 3} \dot F \kappa - \dot F \dot \alpha
       - 2 \left( \ddot F + H \dot F \right) \alpha
       + {1 \over 3} \left( \delta T^{(c)\alpha}_{\;\;\;\;\,\alpha}
       - {\delta F \over F} T^{(c)\alpha}_{\;\;\;\;\,\alpha} \right)
       \Bigg],
   \nonumber \\
   8 \pi G T^0_\alpha
   &=& {1 \over F} \left[ {1 \over a}
       \left(
       {1 \over 2} f_{,X} \dot \phi \delta \phi
       - \delta \dot F + H \delta F
       + \dot F \alpha \right)_{,\alpha}
       + T^{(c)0}_{\;\;\;\;\;\alpha}
       \right],
   \nonumber \\
   8 \pi G \Pi^\alpha_\beta
   &=& {1 \over F} \left[ {1 \over a^2} \left( \nabla^\alpha \nabla_\beta
       - {1 \over 3} \delta^\alpha_\beta \Delta \right)
       \left( \delta F - \dot F \chi \right)
       - \dot F \dot C^\alpha_\beta
       + \delta T^{(c)\alpha}_{\;\;\;\;\,\beta}
       - {1 \over 3} \delta^\alpha_\beta \delta T^{(c)\gamma}_{\;\;\;\;\,\gamma}
       \right],
   \label{pert-fluid-tachyon}
\eea
where we have
\bea
   & & X = - {1 \over 2} \dot \phi^2, \quad
       \delta X = - \dot \phi \delta \dot \phi + \dot \phi^2 \alpha,
\eea
and $R$ and $\delta R$ are given in eqs. (\ref{R},\ref{delta-R}).
The equation of motion gives
\bea
   & & {1 \over a^3} \left( a^3 \dot \phi f_{,X} \right)^\cdot + f_{,\phi}
       = T^{(c)},
   \label{BG-EOM-tachyon} \\
   & & f_{,X} \left[ \delta \ddot \phi
       + \left( 3 H + {\dot f_{,X} \over f_{,X}} \right) \delta \dot \phi
       - {\Delta \over a^2} \delta \phi
       + \dot \phi \left( 3 \dot \varphi - \dot \alpha
       + {\Delta \over a^2} \chi \right) \right]
       + 2 f_{,\phi} \alpha
       + {1 \over a^3} \left( a^3 \dot \phi \delta f_{,X} \right)^\cdot
       + \delta f_{,\phi}
       = \delta T^{(c)},
   \label{pert-EOM-tachyon}
\eea
where
\bea
   & & \delta f = f_{,\phi} \delta \phi + f_{,X} \delta X + f_{,R} \delta R.
\eea
We have located $T^{(c)}_{ab}$ and $T^{(c)}$
terms for later consideration of the string theory correction terms.
In the following we ignore the correction terms.
{}From eqs. (\ref{pert-fluid-GGT},\ref{pert-fluid-decomposition},\ref{Tab})
we have
\bea
   & & 8 \pi G ( \mu + p ) v = - {1 \over aF} \left(
       {1 \over 2} f_{,X} \dot \phi \delta \phi
       - \delta \dot F + H \delta F + \dot F \alpha \right),
   \nonumber \\
   & & 8 \pi G \Pi = { 1\over F} \left( \delta F - \dot F \chi \right), \quad
       8 \pi G \Pi^{(t)\alpha}_{\;\;\;\;\;\beta}
       = - {\dot F \over F} C^{\alpha}_{\beta}.
   \label{pert-fluid-tachyon2}
\eea

We assume $F = F(\phi)$.
{}From eq. (\ref{pert-fluid-tachyon2}) we notice that the uniform-field
gauge differs from the comoving gauge.
Instead of eq. (\ref{Phi-def-MSF}) we introduce a more generalized form
\bea
   & & \Phi \equiv \varphi_{\delta \phi}
       - {K \over a^2} { 2 F \over X f_{,X} + {3 \dot F^2 \over 2 F} } \Psi,
   \label{Phi-def-Tachyon}
\eea
where $\Psi$ is the same as in eq. (\ref{Psi-def}).
{}From eqs. (\ref{G1}-\ref{G3}) and eqs. (\ref{G1},\ref{G3},\ref{G4}),
respectively, we can derive
\bea
   & & \dot \Phi
       = {2 H F + \dot F \over X f_{,X} + {3 \dot F^2 \over 2 F}}
       {c_A^2 \Delta \over a^2} \Psi,
   \label{dot-Phi-eq-Tachyon} \\
   & & { H + {\dot F \over 2F} \over aF}
       \left( {a F \over H + {\dot F \over 2F}} \Psi \right)^\cdot
       = {X f_{,X} + {3 \dot F^2 \over 2F} \over 2 H F + \dot F}
       \Phi,
   \label{dot-Psi-eq-Tachyon}
\eea
where
\bea
   & & c_A^2 \equiv {X f_{,X} + {3 \dot F^2 \over 2 F}
       \over X f_{,X} + 2 X^2 f_{,XX} + {3 \dot F^2 \over 2 F} }
       \left\{ 1 + \left[ 3
       + { X(f_{,X} + 2 X f_{,XX}) + {3 \dot F^2 \over 2F}
       \over Xf_{,X} + {3 \dot F^2 \over 2F} }
       \left( 3 + { {\dot X \over X} - 3 {\dot F \over F} + {\dot E \over E}
       \over H + {\dot F \over 2F} } \right) \right] K \Delta^{-1} \right\},
   \\
   & & E \equiv - {F \over 2X}
       \left( X f_{,X} + {3 \dot F^2 \over 2 F} \right).
   \label{c_A-tachyon}
\eea
Equation (\ref{varphi_chi}) remains valid.
By combining eqs. (\ref{dot-Phi-eq-Tachyon},\ref{dot-Psi-eq-Tachyon}) we have
\bea
   & & {\left( H + {\dot F \over 2F} \right)^2 c_A^2
       \over a^3 \left( X f_{,X} + {3 \dot F^2 \over 2F} \right)}
       \left[ {a^3 \left( X f_{,X} + {3 \dot F^2 \over 2F} \right)
       \over \left( H + {\dot F \over 2F} \right)^2 c_A^2} \dot \Phi
       \right]^\cdot
       = c_A^2 {\Delta \over a^2} \Phi,
   \label{ddot-Phi-eq-Tachyon} \\
   & & {X f_{,X} + {3 \dot F^2 \over 2F} \over
       H F + {1 \over 2} \dot F} \left[
       { \left( H + {\dot F \over 2F} \right)^2 \over
       a \left( X f_{,X} + {3 \dot F^2 \over 2F} \right) }
       \left( {a F \over H + {\dot F \over 2F}} \Psi \right)^\cdot \right]^\cdot
       = c_A^2 {\Delta \over a^2} \Psi.
   \label{ddot-Psi-eq-Tachyon}
\eea
$c_A^2$ differs clearly from $c_s^2 \equiv \dot p / \dot \mu$.
Contrary to the minimally coupled scalar field and
the generalized $f(\phi,R)$ gravity theory
the wave speed is non-trivial even for $K = 0$.
{}For $K = 0$ we have
\bea
   & & c_A^2 = { X f_{,X} + {3 \dot F^2 \over 2 F} \over
       X f_{,X} + 2 X^2 f_{,XX} + {3 \dot F^2 \over 2 F} }.
\eea

{}For the tensor mode, eq. (\ref{GW-eq-GGT}) remains valid exactly.

\subsection{String corrections}
                                               \label{sec:String-corrections}

We consider an action in eq. (\ref{GGT-action}) with the following
additional corrections in the action \cite{HN-string-2000,Cartier-etal-2001}
\bea
   L_{(c)}
   &=& - {1 \over 2} \xi(\phi) \left[
       c_1 R^2_{GB}
       + c_2 G^{ab} \phi_{,a} \phi_{,b}
       + c_3 \Box \phi \phi^{,c} \phi_{,c}
       + c_4 \left( \phi^{,c} \phi_{,c} \right)^2
       \right],
   \label{String-corrections}
\eea
where $R_{GB}^2 \equiv R^{abcd} R_{abcd} - 4 R^{ab} R_{ab} + R^2$.
Corrections to the gravitational field equation and the equation of motion are
\footnote{
          Another forms can be found in eqs. (5-7) of \cite{Cartier-etal-2001}
          with a couple of minor typos in eq. (5):
          these are the fourth term in fourth line
          ($\phi_{\sigma\tau} \rightarrow \phi_{;\sigma\tau}$)
          and the sixth term in fifth line
          ($\xi \delta^\mu_\nu \rightarrow \delta^\mu_\nu$).
          }
\bea
   T^{(c)}_{ab}
   &=&
       - c_1 \Bigg[
       \left( {1 \over 2} R_{GB}^2 g_{ab}
       + 4 R_{ac} R^c_b
       + 4 R^{cd} R_{acbd}
       - 2 R_a^{\;\; cde} R_{bcde}
       - 2 R R_{ab} \right) \xi
   \nonumber \\
   & &
       - 4 \left( \xi^{;cd} R_{acbd}
       - \Box \xi R_{ab}
       + 2 \xi_{;c(b} R^c_{a)}
       - {1 \over 2} \xi_{,a;b} R \right)
       + 2 \left( 2 \xi_{;cd} R^{cd}
       - \Box \xi R \right) g_{ab} \Bigg]
   \nonumber \\
   & &
       - c_2 \Bigg\{
       \xi \left( {1 \over 2} R_{ab} \phi^{,c} \phi_{,c}
       + {1 \over 2} R \phi_{,a} \phi_{,b}
       - 2 R^c_{(a} \phi_{,b)} \phi_{,c} \right)
       - {1 \over 2} \Box \left( \xi \phi_{,a} \phi_{,b} \right)
       - {1 \over 2} \left( \xi \phi^{,c} \phi_{,c} \right)_{,a;b}
       + \left( \xi \phi^{,c} \phi_{,(a} \right)_{;b)c}
   \nonumber \\
   & &
       + {1 \over 2} \left[
       \xi G^{cd} \phi_{,c} \phi_{,d}
       - \left( \xi \phi^{,c} \phi^{,d} \right)_{;cd}
       + \Box \left( \xi \phi^{,c} \phi_{,c} \right)
       \right] g_{ab} \Bigg\}
   \nonumber \\
   & &
       - c_3 \left[
       \left( \xi \phi^{,c} \phi_{,c} \right)_{,(a} \phi_{,b)}
       - \xi \Box \phi \phi_{,a} \phi_{,b}
       - {1 \over 2} \left( \xi \phi^{,c} \phi_{,c} \right)_{,d} \phi^{,d}
       g_{ab} \right]
       - c_4 \xi \phi^{,c} \phi_{,c} \left(
       - 2 \phi_{,a} \phi_{,b}
       + {1 \over 2} \phi^{,d} \phi_{,d} g_{ab} \right),
   \label{GFE-string} \\
   T^{(c)}
   &=& c_1 \xi_{,\phi} R_{GB}^2
       - c_2 G^{ab} \left( \xi_{,\phi} \phi_{,a} \phi_{,b}
       + 2 \xi \phi_{,a;b} \right)
       + c_3 \left[ \xi_{,\phi} \Box \phi \phi^{,a} \phi_{,a}
       + \Box \left( \xi \phi^{,a} \phi_{,a} \right)
       - 2 \left( \xi \Box \phi \phi^{,a} \right)_{;a} \right]
   \nonumber \\
   & &
       + c_4 \left[ \xi_{,\phi} \left( \phi^{,a} \phi_{,a} \right)^2
       - 4 \left( \xi \phi^{,a} \phi^{,b} \phi_{,b} \right)_{;a} \right].
   \label{EOM-string}
\eea
The first line in RHS of eq. (\ref{GFE-string}) vanishes
because we have $\delta \int \sqrt{-g} R_{GB}^2 d^4 x = 0$.

Equations (\ref{BG-eqs},\ref{G7}) remain valid with the same effective fluid
quantities in eqs. (\ref{BG-fluid-GGT},\ref{pert-fluid-GGT})
and the following correction terms.
To the background order we have
\bea
   T^{(c)0}_{\;\;\;\;\; 0}
   &=& - 12 c_1 H \left( H^2 + {K \over a^2} \right) \dot \xi
       + {3 \over 2} c_2 \left( 3 H^2 + {K \over a^2} \right) \xi \dot \phi^2
       - {1 \over 2} c_3 \left( \dot \xi -6 H \xi \right) \dot \phi^3
       + {3 \over 2} c_4 \xi \dot \phi^4,
   \nonumber \\
   T^{(c)\alpha}_{\;\;\;\;\;\alpha}
   &=&
       - 12 c_1 \left[ \left( H^2 + {K \over a^2} \right) \ddot \xi
       + 2 H \left( \dot{H} +H^2 \right) \dot \xi \right]
       + {3 \over 2} c_2 \dot \phi \left[
       \left(2 \dot{H}+ 3 H^2 - {K \over a^2} \right) \xi \dot \phi
       + 4 H \xi \ddot \phi + 2 H \dot \xi \dot \phi \right]
   \nonumber \\
   & &
       + {3 \over 2} c_3 \dot \phi^2 \left( 2 \xi \ddot \phi
       + \dot{\xi}\dot{\phi} \right)
       - {3 \over 2} c_4 \xi \dot{\phi}^4,
   \label{BG-fluid-string} \\
   T^{(c)}
   &=& 24 c_1 \left( \dot{H} + H^2 \right) \left( H^2 + {K \over a^2} \right)
       \xi_{,\phi}
       + 3 c_2 \left[ - \left( H^2+ {K \over a^2} \right)
       \left( \dot{\xi} \dot{\phi} +2\xi\ddot{\phi} \right)
       - 2 H \left( 2 \dot H +3 H^2 + {K \over a^2} \right) \xi \dot \phi
       \right]
   \nonumber \\
   & &
       + c_3 \dot \phi \left[ \ddot \xi \dot \phi
       + 3\dot{\xi}\ddot{\phi}
       - 6 \xi \left( \dot H \dot \phi + 2 H \ddot{\phi}
       + 3 H^2 \dot \phi \right) \right]
       + c_4 \dot{\phi}^2 \left(
       - 3 \dot{\xi} \dot{\phi}-12\xi\ddot{\phi}
       - 12 H \xi \dot{\phi} \right).
   \label{BG-EOM-string-corr}
\eea
To the perturbed order, assuming $K = 0$, we have
\footnote{
          Although eqs. (\ref{pert-fluid-string},\ref{pert-EOM-string-corr})
          were not presented in \cite{Cartier-etal-2001},
          these were derived together with Cyril Cartier while preparing
          \cite{Cartier-etal-2001}.
          }
\bea
   \delta T^{(c)0}_{\;\;\;\;\; 0}
   &=&
       - 4 c_1 H \left[ 3 H^2 \delta \dot \xi
       - 3 H \dot \xi \left( \kappa + H \alpha \right)
       - {\Delta \over a^2} \left( H \delta \xi
       + 2 \dot \xi \varphi \right) \right]
   \nonumber \\
   & &
       + c_2 \dot \phi \left[ {9 \over 2} H^2 \dot \phi \delta \xi
       + 9 H^2 \xi \delta \dot \phi
       - 2 H \xi {\Delta \over a^2} \delta \phi
       - \xi \dot \phi \left( 3 H \kappa
       + 9 H^2 \alpha + {\Delta \over a^2} \varphi \right) \right]
   \nonumber \\
   & &
       - {1 \over 2} c_3 \dot \phi^2
       \left[ 3 \left( \dot \xi - 6 H \xi \right)
       \delta \dot \phi + 2 \xi {\Delta \over a^2} \delta \phi
       + \dot \phi \left( \delta \dot \xi - 6 H \delta \xi \right)
       + 2 \xi \dot \phi \kappa
       - 2 \dot \phi \left( 2 \dot \xi - 9 H \xi \right) \alpha \right]
   \nonumber \\
   & &
       + {3 \over 2} c_4 \dot \phi^3 \left( 4 \xi \delta \dot \phi
       + \dot \phi \delta \xi - 4 \xi \dot \phi \alpha \right),
   \nonumber \\
   \delta T^{(c)\alpha}_{\;\;\;\;\;\alpha}
   &=&
       4 c_1 \Bigg[ - 3 H^2 \delta \ddot \xi
       - 6 H \left( \dot H + H^2 \right) \delta \dot \xi
       + 2 \left( \dot H + H^2 \right) {\Delta \over a^2} \delta \xi
       + 2 H \dot \xi \dot \kappa
   \nonumber \\
   & &
       + 2 \left( H \ddot \xi + \dot H \dot \xi + 3 H^2 \dot \xi \right) \kappa
       + 3 H^2 \dot \xi \dot \alpha
       + 6 H \left( H \ddot \xi + 2 \dot H \dot \xi
       + H^2 \dot \xi \right) \alpha
       + 2 {\Delta \over a^2} \left( \ddot \xi \varphi
       + H \dot \xi \alpha \right)
       \Bigg]
   \nonumber \\
   & &
       + c_2 \Bigg\{
       6 H \xi \dot \phi \delta \ddot \phi
       + 3 \left( 2 \dot H \xi \dot \phi
       + 2 H \xi \ddot \phi + 3 H^2 \xi \dot \phi
       + 2 H \dot \xi \dot \phi \right) \delta \dot \phi
       + 3 H \dot \phi^2 \delta \dot \xi
   \nonumber \\
   & &
       + {3 \over 2} \dot \phi \left( 2 \dot H \dot \phi
       + 4 H \ddot \phi + 3 H^2 \dot \phi \right) \delta \xi
       - \xi \dot \phi^2 \dot \kappa
       - \dot \phi \left( \dot \xi \dot \phi + 2 \xi \ddot \phi
       + 3 H \xi \dot \phi \right) \kappa
       - 6 H \xi \dot \phi^2 \dot \alpha
   \nonumber \\
   & &
       - 9 \dot \phi \left( \dot H \xi \dot \phi
       + 2 H \xi \ddot \phi + H^2 \xi \dot \phi
       + H \dot \xi \dot \phi \right) \alpha
       - {\Delta \over a^2} \left[
       2 \left( \dot \xi \dot \phi + \xi \ddot \phi
       + H \xi \dot \phi \right) \delta \phi
       - \dot \phi^2 \delta \xi
       - \xi \dot \phi^2 \left( \varphi - \alpha \right) \right]
       \Bigg\}
   \nonumber \\
   & &
       + {3 \over 2} c_3 \dot \phi \left[
       2 \xi \dot \phi \delta \ddot \phi
       + \left( 3 \dot \xi \dot \phi + 4 \xi \ddot \phi \right)
       \delta \dot \phi
       + \dot \phi^2 \delta \dot \xi
       + 2 \dot \phi \ddot \phi \delta \xi
       - 2 \xi \dot \phi^2 \dot \alpha
       - 4 \left( \dot \xi \dot \phi^2 + 2 \xi \dot \phi \ddot \phi \right)
       \alpha \right]
   \nonumber \\
   & &
       - {3 \over 2} c_4 \dot \phi^3 \left( 4 \xi \delta \dot \phi
       +\dot \phi \delta \xi -4 \xi \dot \phi \alpha \right),
   \nonumber \\
   \delta T^{(c)0}_{\;\;\;\;\;\alpha}
   &=&
       {1 \over a} \nabla_\alpha \Bigg\{
       {4 \over 3} c_1 H \left[ 3 H \delta \dot \xi
       - 3 H^2 \delta \xi
       - 2 \dot \xi \left( \kappa + {\Delta \over a^2} \chi \right)
       - 3 H \dot \xi \alpha \right]
   \nonumber \\
   & &
       + c_2 \dot \phi \left[
       - 2 H \xi \delta \dot \phi + 3 H^2 \xi \delta \phi
       - H \dot \phi \delta \xi
       + {1 \over 3} \xi \dot \phi \left( \kappa
       + 6 H \alpha + {\Delta \over a^2} \chi \right) \right]
   \nonumber \\
   & &
       - {1 \over 2} c_3 \dot \phi^2 \left[
       2 \xi \delta \dot \phi
       + \left( \dot \xi - 6 H \xi \right) \delta \phi
       + \dot \phi \delta \xi
       - 2 \xi \dot \phi \alpha \right]
       + 2 c_4 \xi \dot \phi^3 \delta \phi
       \Bigg\},
   \nonumber \\
   \delta T^{(c)\alpha}_{\;\;\;\;\;\beta}
       - {1 \over 3} \delta^\alpha_\beta \delta T^{(c)\gamma}_{\;\;\;\;\;\gamma}
   &=&
       {1 \over a^2} \left( \nabla^\alpha \nabla_\beta
       - {1 \over 3} \delta^\alpha_\beta \Delta \right)
       \Bigg\{
       4 c_1 \left[ H \dot \xi \dot \chi
       + \left( H \ddot \xi + \dot H \dot \xi + H^2 \dot \xi \right) \chi
       - \ddot \xi \varphi - H \dot \xi \alpha
       - \left( \dot H + H^2 \right) \delta \xi \right]
   \nonumber \\
   & &
       - {1 \over 2} c_2 \left[
       \dot \phi^2 \delta \xi - 2 \left( \dot \xi \dot \phi + \xi \ddot \phi
       + H \xi \dot \phi \right) \delta \phi
       + \xi \dot \phi^2 \left( \varphi - \alpha + \dot \chi \right)
       + \dot \phi \left( \dot \xi \dot \phi + 2 \xi \ddot \phi
       + H \xi \dot \phi \right) \chi \right]
       \Bigg\}
   \nonumber \\
   & &
       + 4 c_1 \left[ \left( H \dot \xi \dot C^\alpha_\beta \right)^\cdot
       + 3 H^2 \dot \xi \dot C^\alpha_\beta
       - \ddot \xi {\Delta \over a^2} C^\alpha_\beta \right]
       - {1 \over 2} c_2 \left[
       \left( \xi \dot \phi^2 \dot C^\alpha_\beta \right)^\cdot
       + 3 H \dot \phi^2 \xi \dot C^\alpha_\beta
       + \xi \dot \phi^2 {\Delta \over a^2} C^\alpha_\beta \right],
   \label{pert-fluid-string} \\
   \delta T^{(c)}
   &=& c_1 \left( \xi_{,\phi\phi} \bar R^2_{GB} \delta \phi
       + \xi_{,\phi} \delta R^2_{GB} \right)
   \nonumber \\
   & &
       + c_2 \Bigg\{
       - \left[ 6 H^2 \xi \delta \ddot{\phi}
       + 3 H \left(
       4 \dot H \xi + H \dot \xi + 6 H^2 \xi \right) \delta \dot \phi
       - 2 \left( 2 \dot{H} + 3H^2 \right) \xi {\Delta \over a^2}
       \delta \phi \right]
   \nonumber \\
   & &
       - 3 H \left[ H \dot \phi \delta \dot \xi
       + 2 \left( H \ddot \phi + 2 \dot H \dot \phi + 3 H^2 \dot \phi \right)
       \delta \xi \right]
       + 2 {\Delta \over a^2} \left[ 2 H \xi \dot \phi \alpha
       + \left( \dot \xi \dot \phi + 2 \xi \ddot \phi
       + 2 H \xi \dot \phi \right) \varphi \right]
   \nonumber \\
   & &
       +2 \xi \left[2 H \dot\phi\dot\kappa
       + \left( 2 H \ddot \phi + 2 \dot H \dot \phi + 9 H^2 \dot \phi \right)
       \kappa \right]
       + 2 H \dot \xi \dot \phi \left(\kappa + 3 H \alpha \right)
   \nonumber \\
   & &
       + 6 H \xi \left[ H \dot \phi \dot \alpha
       + \left( 2 H \ddot \phi + 4 \dot H \dot \phi + 3 H^2 \dot\phi \right)
       \alpha \right]
       \Bigg\}
   \nonumber \\
   & &
       + c_3 \Bigg\{
       3 \dot \phi \left( \dot \xi - 4 H \xi \right) \delta \ddot \phi
       + \left[ 2 \ddot \xi \dot \phi + 3 \dot{\xi}\ddot{\phi}
       - 12 \xi \left( 3 H^2 \dot \phi
       + \dot H \dot \phi + H \ddot \phi \right) \right] \delta \dot \phi
   \nonumber \\
   & &
       +\left[\dot{\xi}\dot{\phi}+4\xi\left(\ddot{\phi}
       + 2\dot{\phi} H\right)\right]\frac{\Delta}{a^2}
       \delta\phi
   \nonumber \\
   & &
       + \dot{\phi} \left[\dot{\phi} \delta\ddot{\xi}
       + 3\ddot{\phi}\delta\dot{\xi}
       - 6 \left( 2 H \ddot \phi + \dot H \dot \phi + 3 H^2 \dot \phi \right)
       \delta \xi
       -\dot{\phi} \frac{\Delta}{a^2}\delta\xi \right]
       + 2 \xi \dot \phi \left[ \dot \phi \dot \kappa
       + 2 \kappa \left(\ddot \phi + 3 H \dot \phi \right) \right]
   \nonumber \\
   & &
       -4 \dot \phi^2 \dot \alpha \left( \dot \xi - 3 \xi H \right)
       -4 \ddot \xi {\dot \phi}^2 \alpha
       - 12 \dot \xi \dot \phi \ddot \phi \alpha
       + 2 \xi \dot\phi \left[
       9 \left( 2 H \ddot \phi + \dot H \dot \phi + 2 H^2 \dot \phi \right)
       + \dot \phi {\Delta \over a^2} \right] \alpha
       \Bigg\}
   \nonumber \\
   & &
       + c_4 \Bigg\{
       - \dot{\phi} \left[12\xi\dot{\phi}\delta\ddot{\phi}
       + 3 \left( 3 \dot \xi \dot \phi + 8 \xi \ddot \phi
       + 12 H \xi \dot \phi \right)\delta\dot{\phi}
       - 4\xi\dot{\phi}\frac{\Delta}{a^2}\delta\phi\right]
       + 4 \xi \dot \phi^3 \left( \kappa + 3 \dot \alpha \right)
   \nonumber \\
   & &
       -3 \dot{\phi}^2 \left[\dot{\phi}\delta\dot{\xi}
       +4\left(\ddot{\phi} + H \dot \phi \right) \delta \xi \right]
       +12 \alpha \dot{\phi}^2 \left[\dot{\xi}\dot{\phi}
       +\xi\left(4\ddot{\phi} + 3 H \dot \phi \right) \right]
       \Bigg\},
   \label{pert-EOM-string-corr}
\eea
where
\bea
   & & \bar R^{2}_{GB} = 24 H^2 \left( \dot H + H^2 \right), \quad
       \delta R^{2}_{GB} = 4 H^2 \delta R - 16 \dot H
       \left( H \kappa + {\Delta \over a^2} \varphi \right),
   \label{delta-R_GB}
\eea
and $R$ and $\delta R$ are given in eqs. (\ref{R},\ref{delta-R}).

In this and next subsections, for simplicity, we assume $K = 0$;
we believe it is possible, perhaps tedious though,
to derive corresponding equations for general $K$ in similar forms.
We introduce
\bea
   & & \Phi \equiv \varphi_{\delta \phi}, \quad
       \Psi \equiv \varphi_\chi + {\dot F + Q_a \over 2F + Q_b}
       {\delta F_\chi \over \dot F}.
   \label{Psi-def-string}
\eea
{}From eqs. (\ref{G1}-\ref{G3}) and eqs. (\ref{G1},\ref{G3},\ref{G4}),
respectively, we have
\bea
   & & \dot \Phi
       = 2 { \left( H + {\dot F + Q_a \over 2 F + Q_b} \right)
       \left( F + {1 \over 2} Q_b \right)
       \over \omega \dot \phi^2
       + 3 {(\dot F + Q_a)^2 \over 2 F + Q_b} + Q_c }
       {\Delta \over a^2} \Psi,
   \label{dot-Phi-eq-String} \\
   & & { H + {\dot F + Q_a \over 2 F + Q_b} \over a ( F + {1 \over 2} Q_b )}
       \left[ {a (F + {1 \over 2} Q_b) \over
       H + {\dot F + Q_a \over 2 F + Q_b}} \Psi \right]^\cdot
       = {1 \over 2} {
       \omega \dot \phi^2 + 3 {(\dot F + Q_a)^2 \over 2 F + Q_b}
       + Q_c + Q_d + {\dot F + Q_a \over 2F + Q_b} Q_e
       + \left( {\dot F + Q_a \over 2F + Q_b} \right)^2 Q_f
       \over \left( H + {\dot F + Q_a \over 2 F + Q_b} \right)
       \left( F + {1 \over 2} Q_b \right) } \Phi,
   \label{dot-Psi-eq-String}
\eea
where
\bea
   & & Q_{a} \equiv
       - 4 c_1 \dot \xi H^2 + 2 c_2 \xi \dot \phi^2 H + c_3 \xi \dot \phi^3,
   \nonumber \\
   & & Q_{b} \equiv - 8 c_1 \dot \xi H + c_2 \xi \dot \phi^2,
   \nonumber \\
   & & Q_{c} \equiv
       - 3 c_2 \xi \dot \phi^2 H^2
       + 2 c_3 \dot \phi^3 \left( \dot \xi - 3 \xi H \right)
       - 6 c_4 \xi \dot \phi^4,
   \nonumber \\
   & & Q_{d} \equiv
       - 2 c_2 \xi \dot \phi^2 \dot H
       - 2 c_3 \dot \phi^2 \left( \dot \xi \dot \phi
       + \xi \ddot \phi - \xi \dot \phi H \right)
       + 4 c_4 \xi \dot \phi^4,
   \nonumber \\
   & & Q_{e} \equiv
       - 16 c_1 \dot \xi \dot H + 2 c_2 \dot \phi
       \left( \dot \xi \dot \phi + 2 \xi \ddot \phi
       - 2 \xi \dot \phi H \right) - 4 c_3 \xi \dot \phi^3,
   \nonumber \\
   & & Q_{f} \equiv
       8 c_1 \left( \ddot \xi - \dot \xi H \right) + 2 c_2 \xi \dot \phi^2.
   \label{e:Qf_scalar}
\eea
By combining eqs. (\ref{dot-Phi-eq-String},\ref{dot-Psi-eq-String}) we have
\bea
   & & { \left( H + {\dot F + Q_a \over 2 F + Q_b} \right)^2 \over
       a^3 \left( \omega \dot \phi^2 + 3 {(\dot F + Q_a)^2 \over 2 F + Q_b}
       + Q_c \right) }
       \left[ { a^3 \left( \omega \dot \phi^2
       + 3 {(\dot F + Q_a)^2 \over 2 F + Q_b} + Q_c \right) \over
       \left( H + {\dot F + Q_a \over 2 F + Q_b} \right)^2 }
       \dot \Phi \right]^\cdot
       = c_A^2 {\Delta \over a^2} \Phi,
   \label{ddot-Phi-eq-String} \\
   & &
       { \omega \dot \phi^2 + 3 {(\dot F + Q_a)^2 \over 2 F + Q_b}
       + Q_c + Q_d + {\dot F + Q_a \over 2F + Q_b} Q_e
       + \left( {\dot F + Q_a \over 2F + Q_b} \right)^2 Q_f
       \over
       \left( H + {\dot F + Q_a \over 2 F + Q_b} \right)
       \left( F + {1 \over 2} Q_b \right) }
   \nonumber \\
   & & \qquad
       \times
       \left\{ { \left( H + {\dot F + Q_a \over 2 F + Q_b} \right)^2
       \over a \left[ \omega \dot \phi^2 + 3 {(\dot F + Q_a)^2 \over 2 F + Q_b}
       + Q_c + Q_d + {\dot F + Q_a \over 2F + Q_b} Q_e
       + \left( {\dot F + Q_a \over 2F + Q_b} \right)^2 Q_f \right]}
       \left[ {a \left( F + {1 \over 2} Q_b \right) \over
       H + {\dot F + Q_a \over 2 F + Q_b} } \Psi \right]^\cdot \right\}^\cdot
       = c_A^2 {\Delta \over a^2} \Psi,
   \label{ddot-Psi-eq-String}
\eea
where
\bea
   & & c_A^2 = 1 + { Q_d + {\dot F + Q_a \over 2F + Q_b} Q_e
       + \left( {\dot F + Q_a \over 2F + Q_b} \right)^2 Q_f
       \over \omega \dot \phi^2 + 3 {(\dot F + Q_a)^2 \over 2 F + Q_b} + Q_c }.
\eea
A closed form second-order equation in terms of $\Phi$ was derived in
\cite{HN-string-2000,Cartier-etal-2001}.
Using eq. (\ref{delta-F_chi}) we have
\bea
   & & \varphi_\chi
       = {H \Psi + {\dot F + Q_a \over 2 F + Q_b} \varphi_{\delta \phi} \over
       H + {\dot F + Q_a \over 2 F + Q_b}}.
\eea
Notice that the presence of any $c_i$ terms affects $c_A$ in non-trivial ways.

{}For the tensor mode we have \cite{HN-string-2000,Cartier-etal-2001}
\bea
   & & {1 \over a^3 Q_t}
       \left( a^3 Q_t \dot C_{\alpha\beta} \right)^\cdot
       - c_T^2 {\Delta \over a^2} C_{\alpha\beta} = 0,
   \label{GW-eq-String}
\eea
where
\bea
   & & Q_t \equiv F + {1 \over 2} Q_b, \quad
       c_T^2 \equiv 1 - {Q_f \over 2 F + Q_b}.
\eea
This equation is valid for general algebraic function of $f(\phi, R)$.
{}From this wave equation we can read that $c_T$ has a role of the
gravitational wave propagation speed.
$c_T$ is affected by the presence of the $c_1$ and $c_2$ correction terms only.

\subsection{String axionic correction}
                                               \label{sec:String-axion}

We consider an action in eq. (\ref{GGT-action})
with the following additional correction term \cite{Choi-etal-2000}
\bea
   & & L_{(c)} = {1 \over 8} \nu (\phi) R \tilde R,
   \label{String-axion-correction}
\eea
where $R \tilde R \equiv \eta^{abcd} R_{ab}^{\;\;\; ef} R_{cd ef}$ with
$\eta^{abcd}$ a totally antisymmetric Levi-Civita tensor density.
Corrections to the gravitational field equation and the equation of motion are
\bea
   T^{(c)}_{ab}
   &=& \eta_a^{\;\; cde} \left( \nu_{,e;f} R^f_{\;\;bcd}
       - 2 \nu_{,e} R_{bc;d} \right),
   \label{GFE-axion} \\
   T^{(c)}
   &=& - {1 \over 4} \nu_{,\phi} R \tilde R.
   \label{EOM-axion}
\eea
Assuming $K = 0$, the only nonvanishing contribution is
\bea
   & & T^{(c)\alpha}_{\;\;\;\;\;\beta}
       = {1 \over a} \epsilon^{\alpha\gamma\delta}
       \left[ \left( \ddot \nu - H \dot \nu \right)
       \dot C_{\beta\gamma,\delta}
       + \dot \nu D_{\beta\gamma,\delta} \right]
       + ( \alpha \leftrightarrow \beta ),
   \\
   & & T^{(c)} = 0,
\eea
where
\bea
   & & D_{\alpha\beta} \equiv \ddot C_{\alpha\beta}
       + 3 H \dot C_{\alpha\beta} - {\Delta \over a^2} C_{\alpha\beta},
\eea
and we introduced
$\epsilon^{\alpha\beta\gamma} \equiv a^4 \bar \eta^{0\alpha\beta\gamma}$
which is based on $g^{(3)}_{\alpha\beta}$.
Thus, the string axion correction term $\nu R \tilde R$ does not
affect the background equations nor the scalar-type perturbation.
Thus, eqs. (\ref{Psi-def}-\ref{varphi_chi})
in the generalized $f(\phi,R)$ gravity remain valid.
It affects, however, the tensor mode, and
the $(\alpha,\beta)$-component of the field equation, or eqs.
(\ref{GW-eq},\ref{pert-fluid-GGT},\ref{pert-fluid-decomposition}), give
\bea
   & & {1 \over a^3 F} \left( a^3 F \dot C_{\alpha\beta} \right)^\cdot
       - {\Delta \over a^2} C_{\alpha\beta}
       - {2 \over a F} \epsilon_{(\alpha}^{\;\;\;\;\gamma\delta}
       \left[ ( \ddot \nu - H \dot \nu ) \dot C_{\beta)\gamma}
       + \dot \nu D_{\beta)\gamma} \right]_{,\delta}
       = 0.
   \label{GW-eq-String-axion-0}
\eea
This equation is more general than the one derived in \cite{Choi-etal-2000};
it includes our generalized gravity coupling $f(\phi, R)$ in its most
general algebraic form.
We expand \cite{Ford-Parker-1977,H-GW}
\bea
   & & C_{\alpha\beta} ({\bf x}, t)
       \equiv \sqrt{\rm Vol} \int {d^3 k \over ( 2 \pi)^3}
       \sum_\ell e^{(\ell)}_{\alpha\beta} ({\bf k})
       h_{\ell {\bf k}} (t) e^{i {\bf k} \cdot {\bf x}},
\eea
where $e^{(\ell)}_{\alpha\beta}$ is the circular polarization tensor
($\ell = L, R$) with the property
$i k_{\gamma}\epsilon_\alpha^{\;\;\; \gamma\delta} e^{(\ell)}_{\beta\delta}
= k \lambda_\ell e^{(\ell)}_{\alpha\beta}$ ($\lambda_L = - 1$
and $\lambda_R = +1$).
We have
\bea
   & & {1 \over a^3 Q_t} \left( a^3 Q_t \dot h_{\ell {\bf k}} \right)^\cdot
       + {k^2 \over a^2} h_{\ell {\bf k}} = 0,
   \label{GW-eq-String-axion}
\eea
where
\bea
   & & Q_t \equiv F + 2 \lambda_\ell \dot \nu k/a.
\eea
To make this equation similar to the other gravity theories
we may set $\Phi \equiv h_{\ell {\bf k}}$.
We notice that the presence of string-axionic correction term leads
to asymmetric generation and evolution of the two polarization states
of gravitational wave.
However, the wave propagation speed remains $c_A = 1$.

\section{Classical evolution: unified form}
                                               \label{sec:CE}

\subsection{Equations}

All the basic scalar-type perturbation equations considered in
\S \ref{sec:Fluid}-\ref{sec:String-axion} can be written in the following forms
\bea
   & & \dot \Phi = 2 x_1 {\Delta \over a^2} \Psi,
   \label{dot-Phi-eq} \\
   & & {1 \over x_2} \left( x_2 \Psi \right)^\cdot = {1 \over 2} x_3 \Phi.
   \label{dot-Psi-eq}
\eea
Notice that in these forms the normalization of $x_2$ is arbitrary.
In order to be consistent in unified form in the action formulation
in eq. (\ref{perturbed-action}) we fix the normalization in the following way:
we read $x_2$ directly from eqs.
(\ref{dot-Psi-eq-GGT},\ref{dot-Psi-eq-Tachyon},\ref{dot-Psi-eq-String})
whereas from eqs. (\ref{dot-Psi-eq-Fluid},\ref{dot-Psi-eq-MSF}) we read
$x_2 \equiv {1 \over 8 \pi G} {a \over H}$.
In Einstein's gravity limit we have $F = {1 \over 8 \pi G}$.
Introducing
\bea
   & & \bar z \equiv c_A z \equiv \sqrt{a x_2 x_3}, \quad
       c_A \equiv \sqrt{x_1 x_3}; \quad
       \tilde v \equiv z \Phi, \quad
       u \equiv x_2 {1 \over \bar z} \Psi,
   \label{u-v-def}
\eea
we have
\bea
   & & \tilde v = {2 \over c_A \bar z}
       \left( \bar z u \right)^\prime, \quad
       u = {1 \over 2} \Delta^{-1} {z \over c_A}
       \left( {\tilde v \over z} \right)^\prime,
   \label{u-v-eq}
\eea
where a prime indicates a time derivative based on $\eta$
with $dt \equiv a d \eta$.
Thus, we have
\bea
   & & \tilde v^{\prime\prime}
       - \left( c_A^2 \Delta + {z^{\prime\prime} \over z} \right) \tilde v
       = a^2 z \left[ {1 \over a z^2} \left( a z^2 \dot \Phi \right)^\cdot
       - c_A^2 {\Delta \over a^2} \Phi \right] = 0,
   \label{v-eq} \\
   & & u^{\prime\prime} - \left[ c_A^2 \Delta
       + {(1/\bar z)^{\prime\prime} \over (1/\bar z)} \right] u
       = {a^2 x_2 \over \bar z}
       \left\{ {\bar z^2 \over a x_2} \left[ {a \over \bar z^2}
       \left( x_2 \Psi \right)^\cdot \right]^\cdot
       - c_A^2 {\Delta \over a^2} \Psi \right\} = 0.
   \label{u-eq}
\eea
In these forms of the wave equation, $c_A$ has the role of wave speed
of the fluctuating fluid or field and the simultaneously excited metric.
{}For convenience, we summarize various coefficients in Table 1.

\centerline{\bf [[TABLE 1]]}

{}For the tensor mode, using
\bea
   & & z_t \equiv a \sqrt{Q_t}, \quad
       v_t \equiv z_t \Phi,
   \label{z-v-def-GW}
\eea
with $\Phi = C_{\alpha\beta}$ or $h_{\ell {\bf k}}$, we have
\bea
   & & v_t^{\prime\prime}
       - \left( c_T^2 \Delta + {z_t^{\prime\prime} \over z_t} \right) v_t
       = a^2 z_t \left[
       {1 \over a z_t^2} \left( a z_t^2 \dot \Phi \right)^\cdot
       - c_T^2 {\Delta \over a^2} \Phi \right]
       = 0.
   \label{v-eq-GW}
\eea
Thus, it can be absorbed to eq. (\ref{v-eq}) as a unified form.
We summarize various coefficients in the gravitational wave in Table 2.

\centerline{\bf [[TABLE 2]]}

Equations (\ref{v-eq},\ref{v-eq-GW})
in the context of Einstein's gravity are the starting point of
diverse analyses in the context of inflationary structure
generation based on quantum fluctuations.
In this work we have shown that these equations are generally
valid in a wide variety of gravity theories we are considering.

\subsection{Solutions}
                                               \label{sec:solutions}

We have the following general solutions. We introduce the Fourier
transformations of perturbation variables. Since we are considering
linear order perturbation, the equations of the Fourier transformed
variables satisfy the same equations as in the original
configuration space with $\Delta = - k^2$. Thus, we do not
distinguish explicitly the Fourier variables from the original ones.

{\it 1.} In the large-scale limits, with $c_A^2 k^2 \ll z^{\prime\prime}/z$ and
$(1/\bar z)^{\prime\prime}/(1/\bar z)$, we have
\bea
   & & \Phi (k, \eta)
       = {1 \over z} \tilde v
       = C (k) \Bigg\{ 1
       + k^2 \left[ \int^\eta \bar z^2
       \left( \int^\eta {d \eta \over z^2} \right) d \eta
       - \int^\eta \bar z^2 d \eta \int^\eta {d \eta \over z^2} \right] \Bigg\}
       - 2 \tilde d (k) k^2 \int^\eta {d \eta \over z^2},
   \\
   & & \Psi (k, \eta)
       = {\bar z \over x_2} u
       = C (k) {1 \over 2 x_2} \int^\eta \bar z^2 d \eta
       + \tilde d (k) {1 \over x_2} \Bigg\{ 1
       + k^2 \left[ \int^\eta {1 \over z^2}
       \left( \int^\eta \bar z^2 d \eta \right) d \eta
       - \int^\eta \bar z^2 d \eta \int^\eta {d \eta \over z^2} \right]
       \Bigg\}.
\eea
The $C$ ($d$)-mode is relatively growing (decaying) in the expanding phase
of the background world
model\footnote{
         The roles reverse in the collapsing phase.
         In a collapsing phase the $d$-mode rapidly grows and unambiguously
         becomes singular as the background approaches a singularity
         \cite{HN-bounce}.
         }.
Notice that to the leading order in the large-scale expansion
the $C$-mode of $\Phi$ remains constant whereas the one of $\Psi$
changes its behavior according to the background evolution.
Thus, ignoring the transient mode we have
\bea
   & & \Phi (k, \eta) = C (k).
\eea
It is remarkable that
\begin{quote}
``{\sl the constant nature of $\Phi$ in the expanding phase
       is valid independently of changing}
       (i) equation of state $p(\mu)$,
       (ii) field potential $V(\phi)$, and
       (iii) gravity theories $f(\phi,R,X)$, $\omega(\phi)$, $\xi(\phi)$
       and $\nu(\phi)$.''
\end{quote}

{\it 2.} In the small-scale limits, with $c_A^2 k^2 \gg z^{\prime\prime}/z$ and
$(1/\bar z)^{\prime\prime}/(1/\bar z)$, we have
\bea
   & & \tilde v (k, \eta)
       = z \Phi
       = c_{v_1} e^{i c_A k \eta} + c_{v_2} e^{- i c_A k \eta},
   \\
   & & u (k, \eta)
       = {x_2 \over \bar z} \Psi
       = {i \over 2 k} \left(
       - c_{v_1} e^{i c_A k \eta} + c_{v_2} e^{- i c_A k \eta} \right),
\eea
where we assumed $c_A = {\rm constant}$.

Although expressed in general forms, considering the complications in $c_A^2$
for the field and generalized gravities,
these solutions in {\it 1.} and {\it 2.} are properly applicable for $K = 0$.

{\it 3.} We have exact solutions under
\bea
   & & z \propto |\eta|^{q}, \quad
       c_A^2 = {\rm constant}.
   \label{z-condition}
\eea
We have
\bea
   & & {z^{\prime\prime} \over z}
       = {q(q-1) \over \eta^2} \equiv {n \over \eta^2},
\eea
and exact solutions are
\bea
   & & \Phi (k, \eta) = {\sqrt{ \pi |\eta|} \over 2 z}
       \left[ c_1 ({k}) H_\nu^{(1)} (c_A k |\eta|)
       + c_2 ({k}) H_\nu^{(2)} (c_A k |\eta|) \right],
   \label{Phi-exact-sol} \\
   & & \Psi (k, \eta) = - {\sqrt{ \pi |\eta|} \over 2 z}
       {a c_A \over 2 k x_1}
       \left[ c_1 ({k}) H_{\nu - 1}^{(1)} (c_A k |\eta|)
       + c_2 ({k}) H_{\nu - 1}^{(2)} (c_A k |\eta|) \right],
   \label{Psi-exact-sol}
\eea
where
\bea
   & & \nu \equiv {1 \over 2} - q = \sqrt{ n + {1 \over 4}}.
\eea
The normalization is still arbitrary; see eq. (\ref{Phi-mode-sol})
for a normalization assuming the vacuum expectation value.

According to the prescription in eq. (\ref{z-v-def-GW}) the above solution
for $\Phi$ applies to the gravitational wave as well;
see eq. (\ref{P-GW}) for proper normalization.

\section{Quantum generation: unified form}
                                               \label{sec:QG}

\subsection{Quantization}

The perturbed action becomes \cite{action,Mukhanov-1988}
\bea
   \delta^2 S
   &=& {1 \over 2} \int a z^2 \left( \dot \Phi^2
       - c_A^2 {1 \over a^2} \Phi^{,\alpha} \Phi_{,\alpha} \right) dt d^3 x,
   \nonumber \\
   &=& {1 \over 2} \int \left( \tilde v^{\prime 2}
       - c_A^2 \tilde v^{,\alpha} \tilde v_{,\alpha}
       + {z^{\prime\prime} \over z} \tilde v^2 \right)
       d \eta d^3 x.
   \label{perturbed-action}
\eea
The mode expansion is
\bea
   & & \hat \Phi ( {\bf x}, t)
       \equiv \int {d^3 k \over ( 2 \pi)^{3/2} }
       \left[ \hat a_{\bf k} \Phi_{\bf k} (t) e^{i {\bf k}\cdot {\bf x}}
       + \hat a^\dagger _{\bf k} \Phi^*_{\bf k} (t) e^{-i
       {\bf k} \cdot {\bf x}} \right],
   \nonumber \\
   & & [ \hat a_{\bf k} , \hat a_{{\bf k}^\prime} ] = 0, \quad
       [ \hat a^\dagger_{\bf k} , \hat a^\dagger_{{\bf k}^\prime} ] = 0, \quad
       [ \hat a_{\bf k} , \hat a^\dagger_{{\bf k}^\prime} ]
       = \delta^3 ( {\bf k} - {\bf k}^\prime ).
\eea
The conjugate momentum is
$\pi_\Phi \equiv {\partial {\cal L} \over \partial \dot \Phi}
     = a z^2 \dot \Phi$.
The quantization condition
$[ \hat \Phi ({\bf x},t), {\hat \pi}_\Phi ({\bf x}^\prime, t) ]
     = i \delta^3 ({\bf x} - {\bf x}^\prime)$
gives
$[ \hat \Phi ({\bf x},t), \dot {\hat \Phi} ({\bf x}^\prime, t) ]
     = {i \over a z^2} \delta^3 ({\bf x} - {\bf x}^\prime)$
which leads to the Wronskian condition
\bea
   & & \Phi_{\bf k} \dot \Phi_{\bf k}^*
       - \Phi_{\bf k}^* \dot \Phi_{\bf k}
       = {i \over a z^2}.
   \label{Phi-quantization}
\eea
{\it If} the background satisfies eq. (\ref{z-condition}) we have
eqs. (\ref{Phi-exact-sol},\ref{Psi-exact-sol}) as exact solutions.
In terms of the mode function we have \cite{Hwang-MSF,H-Quantum-1997}
\bea
   & & \Phi_{\bf k} (\eta) = {\sqrt{ \pi |\eta|} \over 2 z}
       \left[ c_1 ({k}) H_\nu^{(1)} (c_A k |\eta|)
       + c_2 ({k}) H_\nu^{(2)} (c_A k |\eta|) \right],
   \label{Phi-mode-sol} \\
   & & \Psi_{\bf k} (\eta) = - {\sqrt{ \pi |\eta|} \over 2 z}
       {a c_A \over 2 k x_1}
       \left[ c_1 ({k}) H_{\nu - 1}^{(1)} (c_A k |\eta|)
       + c_2 ({k}) H_{\nu - 1}^{(2)} (c_A k |\eta|) \right],
   \label{Psi-mode-sol}
\eea
where
\bea
   & & |c_2 ({k})|^2 - |c_1 ({k})|^2 = 1,
\eea
which follows from the quantization condition in eq. (\ref{Phi-quantization}).
The two-point function is defined as \cite{Birrell-Davies-1982,Hwang-MSF}
\bea
   & & G_\Phi (x^\prime, x^{\prime\prime})
       \equiv \langle \hat \Phi (x^\prime) \hat \Phi (x^{\prime\prime})
       \rangle_{\rm vac}
       = \int_0^\infty {k^2 d k \over 2 \pi^2}
       j_0 (k|{\bf x}^\prime - {\bf x}^{\prime\prime}|)
       \Phi_{\bf k} (\eta^\prime) \Phi^*_{\bf k} (\eta^{\prime\prime}),
\eea
where $\langle \rangle_{\rm vac}$ is a vacuum expectation value
with $\hat a_{\bf k} | {\rm vac} \rangle \equiv 0$ for every ${\bf k}$;
$x = ({\bf x}, \eta)$.
Using eq. (\ref{Phi-mode-sol}) as the mode function solution, and
assuming the simple vacuum state $c_2 = 1$ and $c_1 = 0$, we have
\cite{Hwang-MSF,H-Quantum-1997}
\bea
   & & G_\Phi (x^\prime, x^{\prime\prime})
       = { \left( {1 \over 4} - \nu^2 \right) \sec{(\pi \nu)} \over
       16 \pi \eta^\prime \eta^{\prime\prime} z^\prime z^{\prime\prime} }
       F \left( {3 \over 2} + \nu; {3 \over 2} - \nu; 2;
       1 + { \Delta \eta^2 - \Delta {\bf x}^2 \over
       4 \eta^\prime \eta^{\prime\prime} } \right),
\eea
which is valid for $\nu < {3 \over 2}$ and
$\Delta \eta^2 - \Delta {\bf x}^2 < 0$;
$\Delta \eta^2 \equiv (\eta^\prime - \eta^{\prime\prime})^2$ and
$\Delta {\bf x}^2 \equiv ({\bf x}^\prime - {\bf x}^{\prime\prime})^2$.

\subsection{Power spectra}
                                               \label{sec:P}

The seed generation process involves three steps.

{\it 1.} We evaluate the power-spectrum based on a vacuum expectation value
introduced as
\bea
   & & {\cal P}_{\hat \Phi} (k,t)
       \equiv {k^3 \over 2 \pi^2} \int
       \langle \hat \Phi ({\bf x} + {\bf r}, t)
       \hat \Phi ({\bf x}, t) \rangle_{\rm vac} e^{-i {\bf k} \cdot {\bf r}}
       d^3 r
       = {k^3 \over 2 \pi^2} | \Phi_k (t) |^2.
   \label{P-vacuum}
\eea
In the large-scale limit using the mode-function solution in
eq. (\ref{Phi-mode-sol}) we have
\bea
   & & {\cal P}^{1/2}_{\hat \Phi} ({\bf k}, \eta)
       = {H \over 2 \pi}
       {1 \over a H |\eta|}
       {\Gamma (\nu) \over \Gamma (3/2)}
       \left( {k |\eta|\over 2} \right)^{3/2 -\nu}
       {1 \over c_A^\nu z/a}.
   \label{P-Phi}
\eea
To include the general vacuum dependence we should multiply
for the scalar-type perturbation
\bea
   & & \big| c_2 ({\bf k}) - c_1 ({\bf k}) \big|,
   \label{vacuum-dependence}
\eea
where
\bea
   & & \big| c_2 ({\bf k}) \big|^2 - \big| c_1 ({\bf k}) \big|^2 = 1.
\eea
{}For $\nu = 0$ we we have additional $2 \ln{(c_A k |\eta|)}$ factor.

{}For the tensor-type perturbation we have
$\hat \Phi = \hat C^\alpha_\beta$ and we need additional
$\sqrt{2}$ factor \cite{H-GW}, with $c_T$ replacing $c_A$, thus
\bea
   & & {\cal P}^{1/2}_{\hat C_{\alpha\beta}} ({\bf k}, \eta)
       = \sqrt{16 \pi G} {H \over 2 \pi} {1 \over a H |\eta|}
       {\Gamma (\nu_t) \over \Gamma (3/2)}
       \left( {k |\eta|\over 2} \right)^{3/2 -\nu_t}
       {1 / \sqrt{8 \pi G} \over c_T^{\nu_t} z_t/a}.
   \label{P-GW}
\eea
To include the general vacuum dependence we should multiply
\bea
   & & \sqrt{ {1 \over 2} \sum_\ell \big| c_{\ell 2} ({\bf k})
       - c_{\ell 1} ({\bf k}) \big|^2 },
   \label{vacuum-dependence-GW}
\eea
where
\bea
   & & \big| c_{\ell 2} ({\bf k}) \big|^2
       - \big| c_{\ell 1} ({\bf k}) \big|^2 = 1.
\eea
$\ell = +, \times$ indicate two polarization states.
{}For $\nu = 0$ we we have additional $2 \ln{(c_A k |\eta|)}$ factor.

In the case of string axionic correction, which will affect only the
gravitational wave, we need to handle the case separately.
We have
\bea
   & & {\cal P}^{1/2}_{\hat C_{\alpha\beta}} ({\bf k}, \eta)
       = \sqrt{16 \pi G} {H \over 2 \pi} {1 \over a H |\eta|}
       {\Gamma (\nu_t) \over \Gamma (3/2)}
       \left( {k |\eta|\over 2} \right)^{3/2 -\nu_t}
       {1 \over c_T^{\nu_t}}
       \sqrt{ {1 \over 2} \sum_\ell
       {1 / 8 \pi G
       \over \left| F + 2 \lambda_\ell \dot \nu {k \over a} \right|}
       \Big| c_{\ell 2} ({\bf k}) - c_{\ell 1} ({\bf k}) \Big|^2 }.
   \label{P-GW-string-axion}
\eea

{\it 2.} In the super-horizon scale we identify \cite{Guth-Pi-1982}
\bea
   & & {\cal P}_{\hat \Phi} \equiv {\cal P}_{\Phi},
\eea
where
\bea
   & & {\cal P}_{\Phi}(k,t)
       \equiv {k^3 \over 2 \pi^2} \int
       \langle \Phi ({\bf x} + {\bf r}, t)
       \Phi ({\bf x}, t) \rangle_{\bf x} e^{-i {\bf k} \cdot {\bf r}}
       d^3 r
       = {k^3 \over 2 \pi^2} | \Phi (k, t) |^2,
   \label{P-spatial}
\eea
is a power-spectrum based on the spatial averaging.
Compare the similarity between eqs. (\ref{P-vacuum},\ref{P-spatial}).

{\it 3.} The growing modes of $\Phi$ are conserved while in the large scale
limit.
Thus, the final classical power spectra of the large-scale structure and the
gravitational wave ${\cal P}_\Phi$ is the same as the ${\cal P}_{\hat \Phi}$
generated from the quantum fluctuations in the early universe.

Spectral indices are defined as
\bea
   n_S - 1, \; n_T
       \equiv {\partial \ln{{\cal P}_\Phi} \over \partial \ln{k}},
   \label{n_ST-def}
\eea
thus
\bea
   {\cal P}_{\Phi} \propto k^{n_S - 1}, \; k^{n_T}.
\eea
Assuming the simplest vacuum state, i.e., $c_2 =1$ and $c_1 = 0$, etc., we have
\bea
   n_S - 1, \; n_T
       = 3 - 2 \nu = 2 + 2 q.
   \label{n_ST}
\eea
In the case of near Harrison-Zel'dovich spectra ($n_S -1 \simeq 0 \simeq n_T$)
the quadrupole anisotropy of the CMB becomes
\bea
   & & \langle a_2^2 \rangle
       = \langle a_2^2 \rangle_S + \langle a_2^2 \rangle_T
       = {\pi \over 75} {\cal P}_{\varphi_{\delta \phi}}
       + 7.74 {1 \over 5} {3 \over 32} {\cal P}_{C_{\alpha\beta}},
   \label{a_2}
\eea
which is valid for $K = 0 = \Lambda$;
for a general situation with nonvanishing $\Lambda$ we need
numerical treatment, see \cite{Knox-1995}.
The ratio between two types of perturbations is
\bea
   & & r_2 \equiv {\langle a_2^2 \rangle_T \over \langle a_2^2 \rangle_S}
       \simeq 3.46 { {\cal P}_{C_{\alpha\beta}} \over
       {\cal P}_{\varphi_{\delta \phi}} }.
\eea
{}From eqs. (\ref{P-Phi},\ref{P-GW}) we have
\bea
   & & r_1 \equiv
       { {\cal P}_{C_{\alpha\beta}} \over {\cal P}_{\varphi_{\delta \phi}} }
       = 2 \left[ \left( {k |\eta| \over 2} \right)^{\nu - \nu_t}
       {\Gamma(\nu_t) \over \Gamma(\nu)}
       {c_A^{\nu-1} \over c_T^{\nu_t}}
       {\bar z \over z_t} \right]^2.
   \label{P-ratio-general}
\eea
Therefore, if the background evolution during the quantum generation
stage satisfies eq. (\ref{z-condition}) we can read the power spectra
(both scalar- and tensor-types) using eqs. (\ref{P-vacuum},\ref{Phi-mode-sol}).
In the large-scale limit we have the power spectra in
eqs. (\ref{P-Phi},\ref{P-GW}).
The spectral indices (slopes) and the ratio of amplitudes
are presented in eqs. (\ref{n_ST},\ref{P-ratio-general}).
The contribution to the quadrupole angular anisotropy can be
estimated using eq. (\ref{a_2}).
We would like to emphasize that all our results in \S \ref{sec:CE} and
\ref{sec:QG} are generally valid in our generalized gravity theories
in unified forms.

The four-year COBE \cite{COBE} data with a $n = 1$ power-law fit give
\bea
   & & Q_{rms-PS|_{n=1}} = 18 \pm 1.6 \mu K, \quad T = 2.725 \pm 0.020 K,
   \nonumber \\
   & & \langle a_2^2 \rangle
       = {4 \pi \over 5} \left( {Q_{rms-PS|_{n=1}} \over T} \right)^2
       \simeq 1.1 \times 10^{-10}.
\eea
The observed quadrupole amplitude is $Q_{rms} = 10^{+7}_{-4} \mu K$
which is lower than the above fitted value.
The first-year WMAP \cite{WMAP} data show even lower value of
$Q_{rms} = 8 \pm 2 \mu K$ and the temperature $T = 2.725 \pm 0.002 K$.
WMAP data also provided a constraint on $r_2$:
$r \equiv 4 {\cal P}_{C_{\alpha\beta}} / {\cal P}_{\varphi_{\delta \phi}}
< 0.90$ with 95\% confidence and $n_S = 0.99 \pm 0.04$, \cite{WMAP}.

\subsection{Slow-roll}
                                               \label{sec:SR}

In the context of $f(\phi, R)$ gravity we have introduced
the following parameters \cite{HN-GGT-1996}
\bea
   & & \epsilon_1 \equiv {\dot H \over H^2}, \quad
       \epsilon_2 \equiv {\ddot \phi \over H \dot \phi}, \quad
       \epsilon_3 \equiv {1 \over 2} {\dot F \over H F}, \quad
       \epsilon_4 \equiv {1 \over 2} {\dot E \over H E}.
   \label{epsilon-def}
\eea
$\epsilon_1$ and $\epsilon_2$ are the slow-roll parameters used
in the minimally coupled scalar field, \cite{SR,SR-second}.
The two additional functional degrees of freedom in $F(\phi)$ and
$\omega(\phi)$ are reflected in $\epsilon_3$ and $\epsilon_4$.
In the context of string correction we have an additional
functional degree of freedom in $\xi(\phi)$.
In order to consider its effect we introduce the following
additional parameters
\bea
   & & \epsilon_5 \equiv {\dot F + Q_a \over H (2 F + Q_b)}, \quad
       \epsilon_6 \equiv {\dot Q_t \over 2 H Q_t},
   \label{epsilon-def2}
\eea
with
\bea
   & & E \equiv {F \over \dot \phi^2} \left( \omega \dot \phi^2
       + 3 {(\dot F + Q_a)^2 \over 2F + Q_b} + Q_c \right).
   \label{E-def2}
\eea
In the $f(\phi, R)$ gravity we have $\epsilon_5 = \epsilon_6 = \epsilon_3$, and
$E$ in eq. (\ref{E-def2}) becomes the one in eq. (\ref{E-def}).
In the case of tachyonic corrections we introduce
\bea
   & & E \equiv - {F \over 2 X} \left( X f_{,X} + 2 X^2 f_{,XX}
       + {3 \dot F^2 \over 2 F} \right),
\eea
and take $\epsilon_5 = \epsilon_6 = \epsilon_3$.

Using the above definitions and eq. (\ref{u-v-def}), assuming $K = 0$,
$z$ and $z_t$ can be written in unified forms
\bea
   & & z \equiv {a \dot \phi/H \over 1 + \epsilon_5} \sqrt{E \over F}, \quad
       z_t \equiv a \sqrt{Q_t}.
\eea
Thus, we can derive
\bea
   {z^{\prime\prime} \over z}
   &=& a^2 \Bigg[
       H^2 \left( 1 - \epsilon_1 + \epsilon_2 - \epsilon_3 + \epsilon_4 \right)
       \left( 2 + \epsilon_2 - \epsilon_3 + \epsilon_4 \right)
       + H \left( - \dot \epsilon_1 + \dot \epsilon_2 - \dot \epsilon_3
       + \dot \epsilon_4 \right)
   \nonumber \\
   & &
       - 2 \left( {3 \over 2} - \epsilon_1 + \epsilon_2 - \epsilon_3
       + \epsilon_4 \right) H {\dot \epsilon_5 \over 1 + \epsilon_5}
       - {\ddot \epsilon_5 \over 1 + \epsilon_5}
       + 2 {\dot \epsilon_5^2 \over ( 1 + \epsilon_5)^2} \Bigg],
   \label{z-over-z-S} \\
   {z_t^{\prime\prime} \over z_t}
   &=& a^2 H^2 \Bigg[ \left( 1 + \epsilon_6 \right)
       \left( 2 + \epsilon_1 + \epsilon_6 \right)
       + {1 \over H} \dot \epsilon_6 \Bigg].
   \label{z-over-z-T}
\eea
Although we have introduced $\epsilon_i$s as extension of slow-roll
parameters, it is far from certain that these parameters
can be properly called the slow-roll parameters.
Thus, it is better to regard $\epsilon_i$s as new definitions of
the fundamental parameters $V(\phi)$, $\omega (\phi)$, $f(\phi, R)$, etc.
If $\dot \epsilon_1 = 0$, we have
\bea
   & & \eta = - {1\over aH} {1\over 1 + \epsilon_1}.
\eea
Thus, for $\dot \epsilon_i = 0$ eqs. (\ref{z-over-z-S},\ref{z-over-z-T}) become
\bea
    {z^{\prime\prime} \over z}
   &=& {1 \over \eta^2} {1\over ( 1 + \epsilon_1 )^2 }
       \left( 1 - \epsilon_1 + \epsilon_2 - \epsilon_3 + \epsilon_4 \right)
       \left( 2 + \epsilon_2 - \epsilon_3 + \epsilon_4 \right)
       \equiv {n_s \over \eta^2},
   \\
   {z_t^{\prime\prime} \over z_t}
   &=& {1 \over \eta^2} {(1 + \epsilon_6) (2 + \epsilon_1 + \epsilon_6)
       \over ( 1 + \epsilon_1 )^2 }
       \equiv {n_t \over \eta^2}.
\eea
The spectral indices become\footnote{
     In order to lift the square root we have assumed
     $3 \ge \epsilon_1 - 2 \epsilon_2 + 2 \epsilon_3 - 2 \epsilon_4$
     and $3 \ge - \epsilon_1 - 2 \epsilon_3$.
     }
\bea
   & & n_S - 1 = 3 - 2 \nu = 3 - \sqrt{4 n_s + 1}
       = 2 {2 \epsilon_1 - \epsilon_2 + \epsilon_3 - \epsilon_4 \over
       1 + \epsilon_1},
   \\
   & & n_T = 3 - 2 \nu_t = 3 - \sqrt{4 n_t + 1}
       = 2 {\epsilon_1 - \epsilon_6 \over 1 + \epsilon_1}.
\eea

If $\epsilon_{1,2,3,4} \ll 1$ we have Harrison-Zel'dovich
spectrum for the scalar-type perturbation, and if
$\epsilon_{1,6} \ll 1$ we have the corresponding one for the
tensor-type perturbation.
In this case eqs. (\ref{P-Phi},\ref{P-GW}) become
\bea
   {\cal P}^{1/2}_{\hat \Phi} ({\bf k}, \eta)
   &=& \left| {H \over 2 \pi} \Big\{ 1 + \epsilon_1
       + \left( 2 \epsilon_1 - \epsilon_2 + \epsilon_3 - \epsilon_4 \right)
       \big[ \ln{(k|\eta|)} - 2 + \ln{2} + \gamma_E \big] \Big\}
       {1 \over c_A^\nu z/a} \right|,
   \label{P-Phi-SR} \\
   {\cal P}^{1/2}_{\hat C_{\alpha\beta}} ({\bf k}, \eta)
   &=& \left| \sqrt{2} {H \over 2 \pi} \Big\{ 1 + \epsilon_1
       + \left( \epsilon_1 - \epsilon_6 \right)
       \big[ \ln{(k|\eta|)} - 2 + \ln{2} + \gamma_E \big] \Big\}
       {1 \over c_T^{\nu_t} z_t/a} \right|,
   \label{P-GW-SR}
\eea
where $\gamma_E = 0.57722$ is the Euler's constant.
Thus,
\bea
   & & n_S - 1 = 2 \left( 2 \epsilon_1 - \epsilon_2 + \epsilon_3
       - \epsilon_4 \right), \quad
       n_T = 2 \left( \epsilon_1 - \epsilon_6 \right),
   \label{SR-indices} \\
   & & r_1 \equiv
       { {\cal P}_{C_{\alpha\beta}} \over {\cal P}_{\varphi_{\delta \phi}} }
       = 2 \left( {c_A^{\nu - 1} \over c_T^{\nu_t}}
       {\bar z \over z_t} \right)^2.
   \label{SR-ratio}
\eea
The spectral indices are generally valid in all gravity theories we are
considering in this work.
In the generalized $f(\phi,R)$ gravity we have
$r_1 = 4 |\epsilon_1 - \epsilon_3| = 2 |n_T|$.
Thus, $r_1 = 4 |\epsilon_1| = 2 |n_T|$ in the minimally coupled scalar field.
The relation $r_1 = 2 |n_T|$ in the minimally coupled scalar field
is known as a `consistency relation'.
We notice that this relation is more generally valid in
generalized $f(\phi,R)$ gravity.
However, in the tachyonic correction we have
$r_1 = 4 |\epsilon_1 - \epsilon_3| c_A = 2 |n_T| c_A$, \cite{HN-Tachyon-2002};
in a simpler case this result was presented in \cite{Garriga-Mukhanov-1999}.
In the case of string correction terms we can derive
\bea
   & & r_1 = 4 \left| \left\{ \epsilon_1 - \epsilon_3
       - {1 \over 4 F} \left[ {1 \over H^2} \left( 2 Q_c + Q_d \right)
       - {1 \over H} Q_e + Q_f \right] \right\}
       {1 \over 1 + {Q_b \over 2 F}}
       \left( {c_A \over c_T} \right)^3 \right|.
\eea
In the string axionic correction term we have
\bea
   & & r_1 = 4 \left| \epsilon_1 - \epsilon_3 \right|
       {1 \over 2} \sum_\ell {1 \over
       \left| 1 + 2 \lambda_\ell {k \over a} {\dot \nu \over F} \right|}.
\eea
Therefore, in the slow-roll limit $\epsilon_i \ll 1$ we have
the tensor-type perturbation suppressed compared with the
scalar-type perturbation.
In the string corrections and the tachyonic correction we have
non-trivial wave propagation speed $c_A$ (see Table 1;
in the string correction case we have non-trivial $c_T$ as well, see Table 2)
and the resulting scalar to tensor ratio $r_1$ could depend on the specific
realization of the background evolution during the quantum generation stage.
The result can be read from eq. (\ref{SR-ratio}) in the slow-roll limit,
or eq. (\ref{P-ratio-general}) in more general situation satisfying only
eq. (\ref{z-condition}).

\section{Discussions}
                                               \label{sec:Discussions}

Considering our own publications on the subject, perhaps it would be
useful to make clear the new points made in this work.
Equations in terms of $\Phi$ and $\Psi$ for the generalized gravity
theories in eqs.
(\ref{dot-Phi-eq-GGT},\ref{dot-Psi-eq-GGT},\ref{dot-Phi-eq-Tachyon},\ref{dot-Psi-eq-Tachyon},\ref{dot-Phi-eq-String},\ref{dot-Psi-eq-String}),
and the corresponding second-order equations in terms of
$\Psi$ in eqs.
(\ref{ddot-Psi-eq-GGT},\ref{ddot-Psi-eq-Tachyon},\ref{ddot-Psi-eq-String})
are new.
We have extended results in \S \ref{sec:GGT} and \ref{sec:Tachyon}
to the situation with general background curvature.
Also, \S \ref{sec:String-axion} is more general by considering
general coupling of the field with gravity.
We stress that the analyses in \S \ref{sec:CE} and \ref{sec:QG}
are made in unified forms applicable to all the generalized gravity theories
we have considered.

Notice that our $f(\phi, R)$ gravity includes $R^2$ gravity as a simple case.
Although $R^2$-term in the action leads to a higher-order gravity theory,
we have shown that we can derive second-order perturbation equations
for the scalar- and tensor-type perturbations in the context of $f(R)$ gravity.
We have investigated the roles of $R^{ab} R_{ab}$ correction term
separately in \cite{NH-Rab};
in four-dimensional spacetime, due to Gauss-Bonnet theorem,
$R^2$ and $R^{ab} R_{ab}$ terms are complete fourth-order contributions
by the pure curvature corrections to quadratic order.
Contrary to $R^2$ gravity, the $R^{ab} R_{ab}$-term leads to fourth-order
differential equations for both scalar- and tensor-type perturbations,
see \cite{NH-Rab}.

No rotational mode (vector-type perturbation) is directly excited by
the presence of generalized forms of the scalar-field and the scalar-curvature;
this is true even in the case of $R^{ab} R_{ab}$ gravity, \cite{NH-Rab}.
This is because the evolution of rotational perturbation
is simply described by the momentum-conservation equation
of the additional fluid part energy-momentum tensor,
$T_{(m)\alpha;b}^{\;\;\;\;\;b} = 0$.

In this work we have considered only a single-component situation.
In the multi-component situation our basic set of equations
in eqs. (\ref{BG-eqs}-\ref{G7}) remains valid with the fluid quantities
interpreted as the sum over the individual fluid quantities including fields.
In order to describe the dynamics of the individual fluid or field component
we additionally need the conservation equations of the
individual energy-momentum tensor or the equation of motion.
Such equations in Einstein's gravity limit and in the generalized $f(\phi,R)$
gravity are presented in \cite{HN-fluids} and \cite{Hwang-1991,HN-CMBR-2002},
respectively.
Extensions to more general situations with $S_{(c)}$ can be
made similarly by considering the multiple fluids and fields
in $T^{(m)}_{ab}$ in eqs. (\ref{GFE-GGT},\ref{GFE-Tachyon}).

We emphasize that most of our equations and analyses made in this work
are independent of the specific scenarios of the universe,
thus can be applied to the spatially homogenous and isotropic
Friedmann world models based on our gravity theories.
In a classic paper on the CMB anisotropies Sachs and Wolfe
\cite{SW-1967} has menthioned that
\begin{quote}
``{\sl the linear perturbations are so surprisingly simple
       that a perturbation analysis accurate to second order may be
       feasible \dots}''
\end{quote}
Considering our unified formulation of perturbations in such wide variety of
gravity theories, the linear perturbations can perhaps be described as
``surprisingly simple'' indeed.
Related to the second part of the statement, an accurate result in
second order perturbation can be found in our recent work in
\cite{NH-NL}.

\section*{Acknowledgments}

HN was supported by grant No. R04-2003-10004-0 from the
Basic Research Program of the Korea Science and Engineering Foundation.
JH was supported by the Korea Research Foundation Grants 2003-015-C00253.


\newpage

\noindent
{\bf Table 1. Scalar-type perturbation:}
We present the coefficients and definitions used in our
unified formulations of the scalar-type perturbation
in \S \ref{sec:CE} and \ref{sec:QG}.
We introduce
$ x_4 \equiv \omega \dot \phi^2 + 3 {(\dot F + Q_a)^2 \over 2 F + Q_b}
  + Q_c + Q_d + {\dot F + Q_a \over 2F + Q_b} Q_e
  + \left( {\dot F + Q_a \over 2F + Q_b} \right)^2 Q_f $.
Except for the string corrections in the last column,
the other situations are valid considering general $K$;
for $c_A^2$ we present result assuming $K = 0$.

\baselineskip=15pt

\noindent
=================================================================
\begin{tabbing}
                              \hskip 0.4cm
\=                            \hskip 0.8cm
 \= Fluid                     \hskip 2.3cm
  \= Field                    \hskip 2.0cm
   \= $f(\phi,R)$ gravity     \hskip 1.0cm
    \= Tachyonic              \hskip 2.0cm
     \= String corrections    \\
--------------------------------------------------------------------------------------------------------------------------------------------------------  \\
$\Phi$ \> $\equiv$
    \> $ \varphi_v - {K \over a^2} {1 \over 4 \pi G ( \mu + p) } \varphi_\chi $
    \> $ \varphi_v - {K \over a^2} {1 \over 4 \pi G \dot \phi^2 } \varphi_\chi $
    \> $ \varphi_{\delta \phi} - {K \over a^2}
         {2 F \over \omega \dot \phi^2 + {3 \dot F^2 \over 2 F}} \Psi $
    \> $ \varphi_{\delta \phi}
         - {K \over a^2} { 2 F \over X f_{,X} + {3 \dot F^2 \over 2 F} } \Psi $
    \> $ \varphi_{\delta \phi} $
    \\
--------------------------------------------------------------------------------------------------------------------------------------------------------  \\
$\Psi$ \> $\equiv$
    \> $ \varphi_\chi $
    \> $ \varphi_\chi $
    \> $ \varphi_\chi + {\delta F_\chi \over 2 F} $
    \> $ \varphi_\chi + {\delta F_\chi \over 2 F} $
    \> $ \varphi_\chi + {\dot F + Q_a \over 2F + Q_b}
         {\delta F_\chi \over \dot F} $
    \\
--------------------------------------------------------------------------------------------------------------------------------------------------------  \\
$x_1$ \> $\equiv$
    \> $ {H \over 8 \pi G ( \mu + p) } c_s^2 $
    \> $ {H \over 8 \pi G \dot \phi^2} c_A^2 $
    \> $ {H F + {1 \over 2} \dot F \over \omega \dot \phi^2
          + {3 \dot F^2 \over 2 F}} c_A^2 $
    \> $ {H F + {1 \over 2} \dot F \over
         X f_{,X} + {3 \dot F^2 \over 2 F}}
         c_A^2 $
    \> $ { \left( H + {\dot F + Q_a \over 2 F + Q_b} \right)
         \left( F + {1 \over 2} Q_b \right)
         \over \omega \dot \phi^2
         + 3 {(\dot F + Q_a)^2 \over 2 F + Q_b} + Q_c } $
    \\
--------------------------------------------------------------------------------------------------------------------------------------------------------  \\
$x_2$ \> $\equiv$
    \> $ {1 \over 8 \pi G} {a \over H} $
    \> $ {1 \over 8 \pi G} {a \over H} $
    \> $ {a F \over H + {\dot F \over 2F}} $
    \> $ {a F \over H + {\dot F \over 2F}} $
    \> $ {a (F + {1 \over 2} Q_b) \over
       H + {\dot F + Q_a \over 2 F + Q_b}} $
    \\
--------------------------------------------------------------------------------------------------------------------------------------------------------  \\
$x_3$ \> $\equiv$
    \> $ 8 \pi G {\mu + p \over H} $
    \> $ 8 \pi G {\dot \phi^2 \over H} $
    \> $ {\omega \dot \phi^2 + {3 \dot F^2 \over 2F}
         \over H F + {1 \over 2} \dot F} $
    \> $ {X f_{,X} + {3 \dot F^2 \over 2F} \over H F + {1 \over 2} \dot F} $
    \> $ {1 \over \left( H + {\dot F + Q_a \over 2 F + Q_b} \right)
         \left( F + {1 \over 2} Q_b \right)} x_4 $
    \\
--------------------------------------------------------------------------------------------------------------------------------------------------------  \\
$c_A^2$ \> $\equiv$
    \> $ c_s^2 \left( \equiv {\dot p \over \dot \mu} \right) $
    \> $ 1 $
    \> $ 1 $
    \> $ { X f_{,X} + {3 \dot F^2 \over 2 F} \over
         X f_{,X} + 2 X^2 f_{,XX} + {3 \dot F^2 \over 2 F} } $
    \> $ { x_4 \over \omega \dot \phi^2
         + 3 {(\dot F + Q_a)^2 \over 2 F + Q_b} + Q_c } $
    \\
--------------------------------------------------------------------------------------------------------------------------------------------------------  \\
$\bar z$ \> $\equiv$
    \> $ {a \over H} \sqrt{\mu + p} $
    \> $ {a \over H} \dot \phi $
    \> $ {a \over H + {\dot F \over 2F}}
         \sqrt{ \omega \dot \phi^2 + {3 \dot F^2 \over 2 F} } $
    \> $ {a \over H + {\dot F \over 2F}}
         \sqrt{ X f_{,X} + {3 \dot F^2 \over 2 F} } $
    \> $ {a \over H + {\dot F + Q_a \over 2 F + Q_b} } \sqrt{x_4} $
    \\
--------------------------------------------------------------------------------------------------------------------------------------------------------  \\
$u$ \> $\equiv$
    \> $ {1 \over 8 \pi G \sqrt{\mu + p}} \Psi $
    \> $ {1 \over 8 \pi G \dot \phi} \Psi $
    \> $ {F \over \sqrt{ \omega \dot \phi^2 + {3 \dot F^2 \over 2 F} } } \Psi $
    \> $ {F \over \sqrt{ X f_{,X} + {3 \dot F^2 \over 2 F} } } \Psi $
    \> $ {F + {1 \over 2} Q_b \over \sqrt{x_4}} \Psi $
    \\
--------------------------------------------------------------------------------------------------------------------------------------------------------  \\
\end{tabbing}

\noindent
{\bf Table 2. Tensor-type perturbation:}
Continuation of Table 1 for the tensor-type perturbation (gravitational wave).
In the cases of the string corrections and the string axion we assume $K = 0$;
for $c_T^2$ we present result assuming $K = 0$.

\noindent
=================================================================
\begin{tabbing}
                                    \hskip 0.4cm
\=                                  \hskip 1.8cm
 \= Fluid, Field                    \hskip 1.2cm
  \= $f(\phi,R)$ gravity, Tachyonic \hskip 1.0cm
   \= String corrections            \hskip 1.2cm
    \= String axion                 \\
--------------------------------------------------------------------------------------------------------------------------------------------------------  \\
$z_t$ \> $\equiv$
    \> $ a {1 \over \sqrt{8 \pi G}} $
    \> $ a \sqrt{F} $
    \> $ a \sqrt{F + {1 \over 2} Q_b} $
    \> $ a \sqrt{F + 2 \lambda_\ell \dot \nu k/a} $
    \\
--------------------------------------------------------------------------------------------------------------------------------------------------------  \\
$c_T^2$ \> $\equiv$
    \> $ 1 $
    \> $ 1 $
    \> $ 1 + {Q_f \over 2 F + Q_b} $
    \> $ 1 $
    \\
--------------------------------------------------------------------------------------------------------------------------------------------------------  \\
\end{tabbing}

\end{document}